\documentclass[journal]{IEEEtran}
\pdfoutput=1
\usepackage{graphicx}
\usepackage{amsfonts}
\usepackage{amssymb}
\usepackage[cmex10]{amsmath}
\begin{document}
\title{Minimal Compression of a Radio-Frequency Pulse} 
\author{W.~J.~Szajnowski~\IEEEmembership{}
        {}% <-this % stops a space
\thanks{W. J. Szajnowski is a Visiting Professor at %VSSP 
	Centre for Vision, Speech and Signal Processing, University of Surrey, Guildford GU2 7XD, U.K., e-mail: w.j.szajnowski@surrey.ac.uk}% <-this % stops a space
\thanks{}
}

\markboth{ }%
{Shell \MakeLowercase{\textit{et al.}}: Bare Demo of IEEEtran.cls for Journals}

\maketitle

\begin{abstract}
\boldmath
 Quadrature amplitude modulation (QAM) and a complementary representation of a causal waveform have been used to develop a sidelobe-free pulse-compression technique. Envelopes of radio-frequency (RF) pulses under study include both unimodal (Laplacian, Gaussian, rectangular) and bimodal (Hermite-Gaussian) shapes. Although the achievable compression gain is small ($\sim\mspace{-5mu}2$), the generation and phase-invariant correlation processing of such created compressible pulses can be accomplished with the use of low-complexity systems. 
 
 \end{abstract}

\begin{IEEEkeywords}
 Pulse compression, quadrature amplitude modulation, phase-invariant processing, Kramers-Kronig relations 
\end{IEEEkeywords}

\IEEEpeerreviewmaketitle

\section{Introduction}
\IEEEPARstart{I}{n} the time domain, a linear time-invariant system is completely characterised by its impulse response, i.e. a signal observed at the output of a system under test when its input is excited by the Dirac delta (or impulse) function. 

When correlation analysis is used to determine the impulse response of a system, the autocorrelation function of a waveform employed to probe the system should assume the form of a short pulse having a pronounced peak and zero or low values outside this peak, i.e. an approximation of the Dirac delta function [1].

In a broad class of applications, including autonomous navigation, radar or microwave imaging, a radio-frequency (RF) waveform is used as a probing signal. When the response of an interrogated system is to be detected in additive wideband noise, an optimal receiver will usually incorporate a matched filter or a correlator [1]--[3]. 

The output waveform of an optimal receiver comprises two components: a representation of the autocorrelation function of a probing RF waveform and a suitably attenuated input noise. Therefore, the shape of the probing waveform and the shape of its autocorrelation function are both important factors affecting waveform design procedures. 

In radar systems, a phase or frequency modulation technique, referred to as {\emph {pulse compression}}, is often exploited to expand the frequency spectrum of a probing waveform [2]. While the resulting autocorrelation function has a reduced width, it may also exhibit, in addition to its main narrow peak, some local maxima (sidelobes) that will create 'ghost' responses and may  obscure smaller objects.

In many practical applications, such as those involving low-power autonomous systems or distributed sensor networks, it is required that both the generation and correlation processing of proposed compressible waveforms should be implemented with the use of standard low-complexity analogue or digital building blocks.

\section{Preliminary Considerations}
A pulse $s(t)$ of a radio-frequency (RF) carrier is usually represented 
by a bandpass model [2],\,[3],
\begin{equation}
	s(t) \,=\, \mu(t) \cos \mspace{-2mu}
	\left[   \mspace{1mu}           \omega_0\mspace{1mu}t \pm \psi(t) \right]
\end{equation}
where $\mu(t)$, $\psi(t)$ and $\omega_0$ are, respectively, the {\emph {natural}} 
pulse envelope, the phase modulating function and the angular frequency. 

It is assumed that the functions, $\mu(t)$ and $\psi(t)$, are both slow varying with respect to the carrier oscillations. Accordingly, the extent of the Fourier transforms of $\mu(t)$ and $\psi(t)$ is assumed to be limited to the interval $(-\omega_0/2,\omega_0/2)$.

Fig.\,1 shows examples of some unimodal symmetric shapes$\mspace{1mu}${\footnote {\,The Fourier transforms of these shapes are not band limited; in practice, the carrier frequency $\omega_0$ must be sufficiently high to justify the assumption.}} of natural pulse envelopes:

\vspace{0.07cm} 
\noindent
1. Gaussian, $\exp(-\mspace{1mu}t^2)$;\\ 
2. Laplacian, $\exp(-|t|)$;\\
3. 'soft' rectangular,
$\{\tanh[\pi (t\mspace{-2mu}+\mspace{-2mu}1)]\mspace{-1mu}
-\mspace{-1mu}\tanh[\pi(t\mspace{-2mu}-\mspace{-2mu}1)]\}\mspace{-2mu} /2$.

\vspace{0.07cm} 
\noindent
With respect to the {\emph {mesokurtic}} shape of a Gaussian envelope, a Laplacian envelope is {\emph {leptokurtic}}, whereas a rectangular one is {\emph {platykurtic}}.

\begin{figure}[] %[b]
	\centering
	\includegraphics[width=7.5cm]{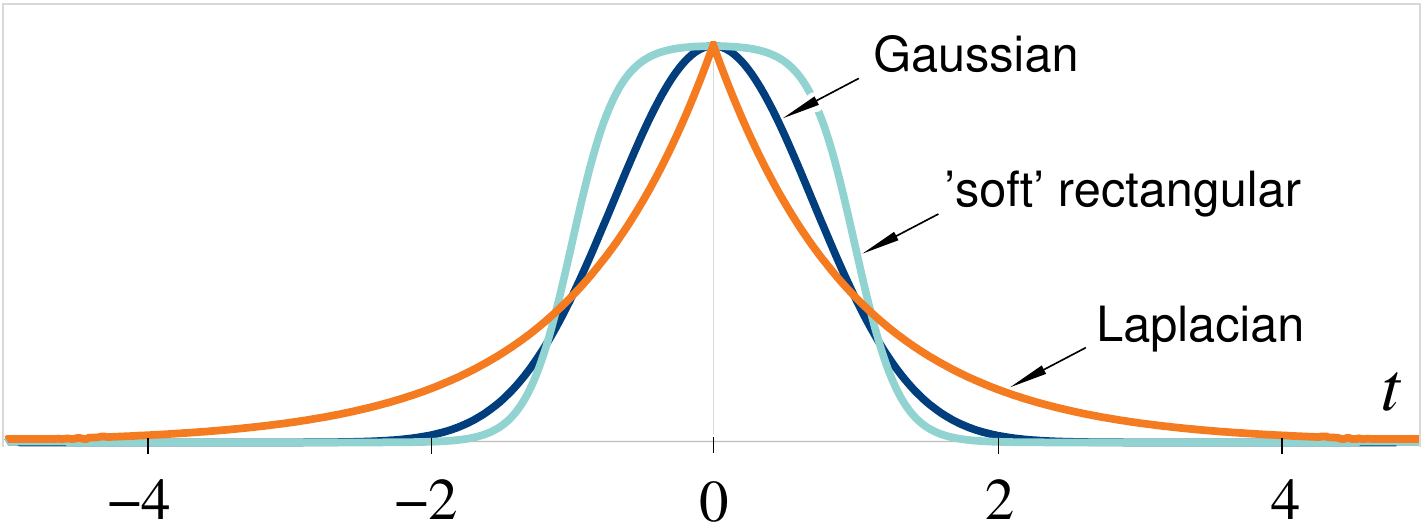}
	\vspace{-0.3cm}
	\caption{Symmetric pulse envelopes having different shapes: Gaussian, Laplacian and 'soft' rectangular.}
\end{figure}

A 'soft' rectangular envelope can be expressed as
\begin{equation}
	\mu(t) \,=\, 
	\{\tanh\mspace{1mu}[\gamma\mspace{1mu} (t\mspace{-1mu}+\mspace{-1mu}\kappa)]
	-\tanh\mspace{1mu}[\gamma\mspace{1mu}(t\mspace{-1mu}-\mspace{-1mu}\kappa)]
	\mspace{1mu}\}\mspace{-1mu} /2
\end{equation}
where $-\kappa$ and $+\kappa$ are nominal positions of its edges, and the 
 shape parameter $\gamma$ characterises their steepness.
When $\gamma$ approaches infinity, $\tanh(\gamma\mspace{1mu} t)\to \mathrm{sgn}(t)$, 
\noindent 
where
\begin{equation}
	\mathrm{sgn}(\zeta)\, \triangleq \,
	\left\{
	\begin{array}{r@{\,\,\quad}l}\!\!+1, & {\zeta>0}\\
		\!\!0, & {\zeta=0}\\
		\!\!-1, & {\zeta<0}\mspace{1mu}
	\end{array} \right . \nonumber 
\end{equation}
and (2) represents a rectangular envelope,\\ 

\vspace{-0.2cm}
{\centerline {$\mu(t) \,=\, 
[\mspace{1mu}\mathrm{sgn}\mspace{1mu} (t\mspace{-1mu}+\mspace{-1mu}\kappa) - \mathrm{sgn}\mspace{1mu}(t\mspace{-1mu}-\mspace{-1mu}\kappa)]/2\mspace{1mu}.$}}

\subsection{Causal (One-Sided) Waveforms} 
Any even (or odd) function of time can be represented by a 'one-sided' function, since the notion of symmetry (or antisymmetry) is associated with inherent redundancy of shape information. 

\begin{figure}[] %[b]
	\centering
	\includegraphics[width=7.5cm]{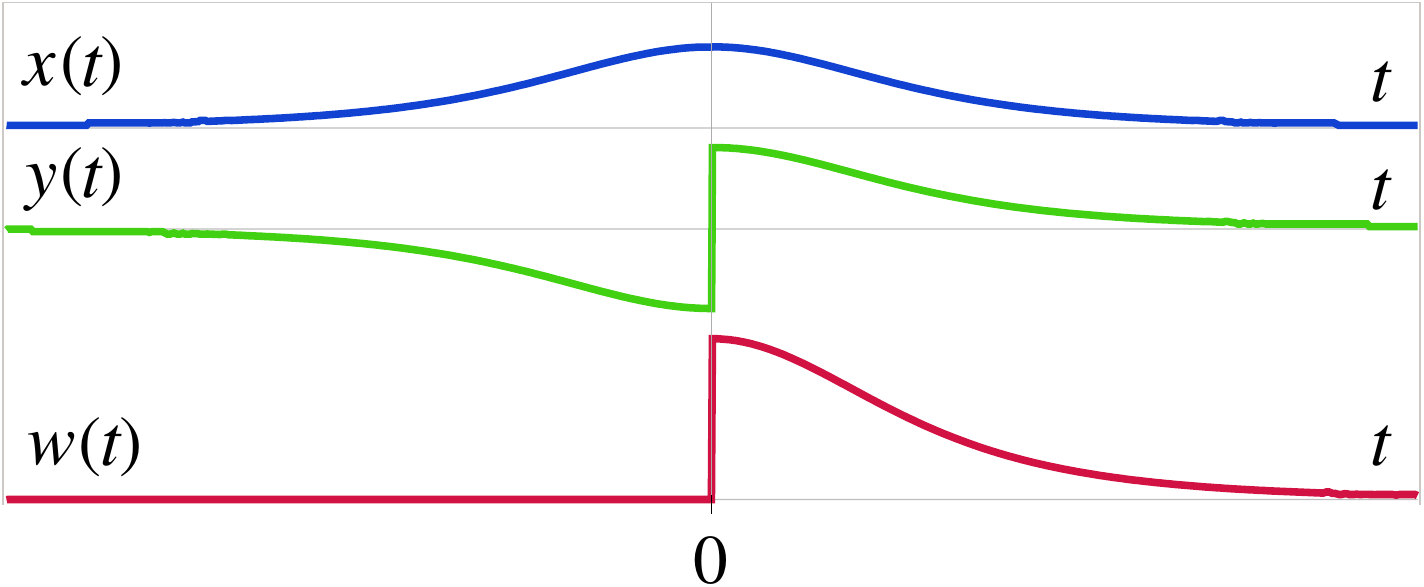}
	\vspace{-0.3cm}
	\caption{Symmetric pulse waveform $x(t)$, its  antisymmetric replica $y(t)$ and a resulting one-sided waveform $w(t)$.}
\end{figure}

A symmetric pulse waveform $x(t)$ can be used to form its antisymmetric replica, $y(t)$, as 
follows$\mspace{1mu}${\footnote {\,Such a transformation will also produce a symmetric replica of an antisymmetric pulse.}}
\begin{equation}
	y(t) = x(t) \mspace{2mu}\mathrm{sgn}(t)\mspace{1mu}.
\end{equation}
The energy, $E_y$, of the antisymmetric replica,
\begin{equation*}
	E_y \mspace{3mu} \triangleq \int_{-\infty}^{\infty}\!\!
	y^2(t) \, dt \mspace{3mu}
	= \int_{-\infty}^{\infty}\!\!x^2(t) \mspace{2mu}[\mathrm{sgn}(t)]^2\mspace{1mu}
	dt\mspace{3mu} = E_x
\end{equation*}
is the same as the energy, $E_x$, of the underlying symmetric waveform $x(t)$.

Consider the sum of the waveforms, $x(t)$ and $y(t)$,  
\begin{eqnarray}
	w(t) \!\!\!&=&\mspace{-8mu} x(t)+y(t) \,=\, x(t)[1+ \mathrm{sgn}(t)] \nonumber \\
	&=&\mspace{-8mu}2\mspace{1mu} \mathsf{H}(t)\mspace{1mu} x(t) \, = \,	
	{ 	\left\{
		\begin{array}{l@{\,\,\quad}l}\!\!2 x(t), & {t>0}\\
			\!\!x(0), & {t=0}\\
			\!\!0, & {t<0} 
		\end{array} \right . }
\end{eqnarray}
where $\mathsf{H}(\cdot)$ is the unit step (Heaviside) function, defined by
\begin{equation}
	\mathsf{H}(\zeta)\, \triangleq \,
	\left\{
	\begin{array}{r@{\,\,\quad}l}\!\!+1, & {\zeta>0}\\
		\!\!1/2, & {\zeta=0}\\
		\!\!0, & {\zeta<0}\mspace{1mu}.
	\end{array} \right . \nonumber 
\end{equation}
The resulting waveform $w(t)$, shown in Fig.\,2, is a {\emph {causal}} waveform 
that vanishes$\mspace{1mu}${\footnote{\,In a similar fashion, in a single-sideband (SSB) communication system, a double-sided frequency spectrum of a waveform is converted into a one-sided spectrum.}} for $t < 0$.

The waveform $w(t)$ may then be regarded as a {\emph {compressive}} (non-redundant) representation of the underlying symmetric pulse waveform $x(t)$. In the following, this specific form of pulse compression will be referred to as {\emph {minimal}} compression.

\subsection{Quadrature Modulation}
 An RF pulse (1) can also be expressed in a {\emph {canonical}} form
\begin{eqnarray}
	s(t) \mspace{-10mu}&=&\mspace{-10mu} 
	{\underbrace{\mspace{2mu}\mu(t) \cos[\psi(t)]\mspace{1mu}}
		_{\mspace{7mu}x(t)}} \cos(\omega_0\mspace{1mu}t) \mp
	{\underbrace{\mspace{2mu}\mu(t) \sin[\psi(t)]\mspace{1mu}}
		_{\mspace{7mu}y(t)}} \sin(\omega_0\mspace{1mu}t) \nonumber \\
	\mspace{-10mu}&=&\mspace{-9mu}
	x(t)\cos(\omega_0\mspace{1mu}t) \mp
	y(t) \sin(\omega_0\mspace{1mu}t)
\end{eqnarray}
where the {\emph {in-phase}} pulse component, $x(t)$, and the {\emph {quadrature}} pulse component, $y(t)$, are defined by
\begin{equation}
	x(t) \, \triangleq \, \mu(t) \cos[\psi(t)],  \qquad
	y(t) \, \triangleq \, \mu(t) \sin[\psi(t)] . \nonumber
\end{equation}

\begin{figure}[] %[b]
	\centering
	\includegraphics[width=6.5cm]{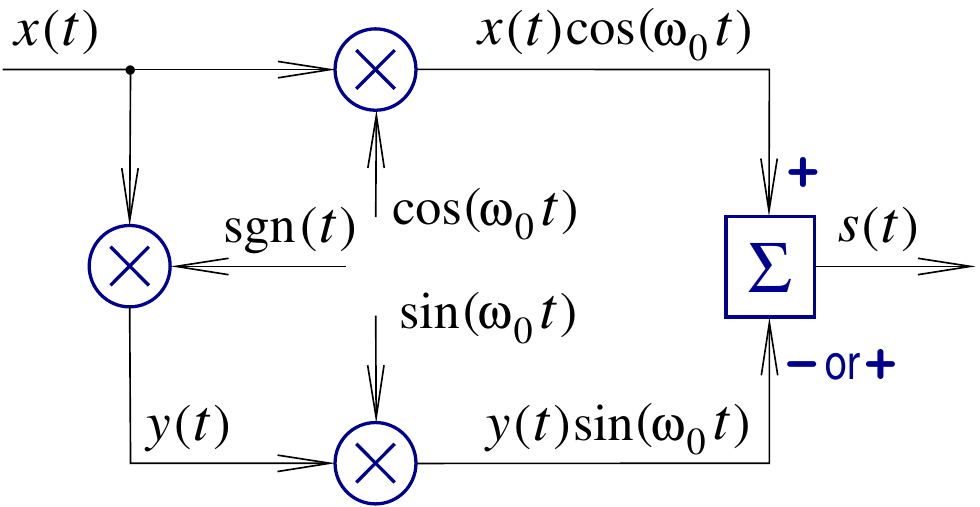}
	\vspace{-0.2cm}
	\caption{Formation of a minimally compressible pulse of a harmonic carrier wave.}
\end{figure}

Consequently, an RF pulse (1) can be regarded as a result of modulating two orthogonal components, $\cos(\omega_0\mspace{1mu} t)$ and $\sin(\omega_0\mspace{1mu} t)$, of its carrier by two pulse waveforms, $x(t)$ and $y(t)$. In communication systems, the form (5) is referred to as quadrature amplitude modulation (QAM).

Fig. 3 illustrates the concept of formation of a minimally compressible pulse with the use of a quadrature modulator. 

In this arrangement, a causal waveform $w(t)$ is {\emph {virtual}}, i.e. not reconstructed in any explicit way. However, its two components, $x(t)$ and $y(t)$, are used to modulate the two orthogonal components of an RF carrier. Therefore, the output composite waveform, $s(t)$, can be regarded as an implicit representation of the causal waveform $w(t)$.

\subsection{Problem Formulation}
It is of theoretical and practical interest to exploit the structure (4) of a causal waveform in conjunction with the canonical form (5) of quadrature modulation to construct an RF pulse having a prescribed symmetric envelope and exhibiting improved time resolution. 

In many practical applications, such as radar, the phase offset of a received RF pulse is unknown and cannot be estimated in any reliable manner. Therefore, it is also of interest to develop a phase-invariant signal processor capable of extracting from a received waveform a suitable representation of the autocorrelation function of a transmitted pulse.

\section{A Minimally Compressible Pulse Waveform}
A {\emph {complex envelope}}, $z(t)$, of an RF pulse (5) is defined as follows [3]
\begin{equation}
	z(t)\, \triangleq \,x(t) + j y(t) 
\end{equation}
where $j^2\!=\!-1$.
The complex envelope can then be used to represent an RF pulse as
\begin{equation}
	s(t)\mspace{1mu} = \mspace{1mu}\mathcal{R}e\{z(t) \exp\mspace{1mu}(\pm j\mspace{1mu}\omega_0 t)\}
\end{equation}
where $\mathcal{R}e\{\cdot\}$ denotes the real part.

The natural envelope $\mu(t)$ equals the magnitude of the complex envelope,
\begin{equation}
	\mu(t) \mspace{1mu} = \mspace{1mu} |z(t)| \mspace{1mu} = \mspace{1mu} \sqrt{x^2(t) + y^2(t)} .
\end{equation}
and, in accordance with (1) and (5), the phase modulating function $\psi(t)$ can be determined from
\begin{eqnarray}
	\psi(t) \mspace{1mu} = \mspace{2mu} \arctan\mspace{1mu}[\mspace{1mu}{y(t)/x(t)}] .
	%\nonumber \\
	     %   \mspace{-8mu}&=&\mspace{-8mu}  (\pi/4)\mspace{1mu}\mathrm{sgn}(t).
\end{eqnarray}

A complex envelope $z(t)$ can be regarded as a {\emph {complex}} representation of a causal pulse waveform $w(t)$. Such representation facilitates the development of matched filters and correlators intended for processing of RF pulses corrupted by additive wideband noise.

\subsection{Kramers-Kronig Relations: A Causal Waveform}
In the frequency domain, a causal waveform (4) is represented by its Fourier transform$\mspace{1mu}${\footnote {\,The expression (10) is also the Fourier transform of a conjugate envelope
 $z^{\!*}(t) \triangleq \mspace{1mu} x(t)\mspace{-2mu}-\mspace{-2mu}jy(t)$.}}, 
\begin{eqnarray}
	W(\omega)\, = \,\mathcal{F}\{w(t)\} \mspace{-9mu}&\triangleq&\mspace{-9mu}
	\int_{-\infty}^{\infty}\mspace{-5mu}
	w(t) \exp(-j \omega\mspace{1mu} t) \mspace{1mu} dt \nonumber \\
		\mspace{-9mu}&=&\mspace{-7mu}	X(\omega) + j Y(\omega)
\end{eqnarray}
where $\mathcal{F}(\cdot)$ denotes the Fourier transform and $\omega$ is the angular frequency. The real and imaginary parts of $W(\omega)$ are	given by
\begin{equation*}
	X(\omega) = 2\mspace{-4mu}\int_{0}^{\infty}\mspace{-8mu}
	x(t) \cos(\omega\mspace{1mu} t) \mspace{1mu} dt, \,\,   
	Y(\omega) = -2\mspace{-4mu}\int_{0}^{\infty}\mspace{-8mu}
	y(t) \sin(\omega\mspace{1mu} t) \mspace{1mu} dt .
\end{equation*}
In the considered case, $y(t) = x(t)\mspace{1mu}\mathrm{sgn}(t)$; since multiplication of two functions of time is equivalent to convolving, in the frequency domain, their respective Fourier transforms, the imaginary part $Y(\omega)$ can be expressed as
\begin{equation}
	Y(\omega)\, = \, \mathcal{F}\{x(t)\} \star \mathcal{F}\{\mathrm{sgn}(t)\}
\end{equation}
where $\star$ denote convolution.

Since [5]
\begin{equation}
	\mathrm{sgn}(t)	\stackrel{{}\mspace{4mu}\mathcal{F}}{\iff}2/(j\mspace{1mu}\omega)
\end{equation}
the imaginary part can be determined from
\begin{eqnarray}
	Y(\omega)\, = \, X(\omega) \star (-2/\omega)
	\mspace{-9mu}&=&\mspace{-9mu} -\mathcal{H}\{X(\omega)\} \nonumber \\
	\mspace{-9mu}&=&\mspace{-9mu} {\mathcal{H}}^{-1}\{X(\omega)\} 
\end{eqnarray}
where $\mathcal{H}(\cdot)$ denotes the Hilbert transform [4], [8].
The above result can be expressed symbolically as
\begin{equation}
	X(\omega)\,\,\, {\Large{
		{\substack{\scriptscriptstyle{{\mathcal{H}}}\\
	{{\rightleftarrows}}\\
		\scriptscriptstyle{-\mathcal{H}}} 
	}} 	\, -Y(\omega)}.
\end{equation}
Therefore, $W(\omega)$ is an analytic function. 

The fact that a causal waveform has an analytic Fourier spectrum is the essence of the Kramers-Kronig relation for causal waveforms [4], [6].

\begin{figure}[t] %[b]
	\centering
	\includegraphics[width=8.5cm]{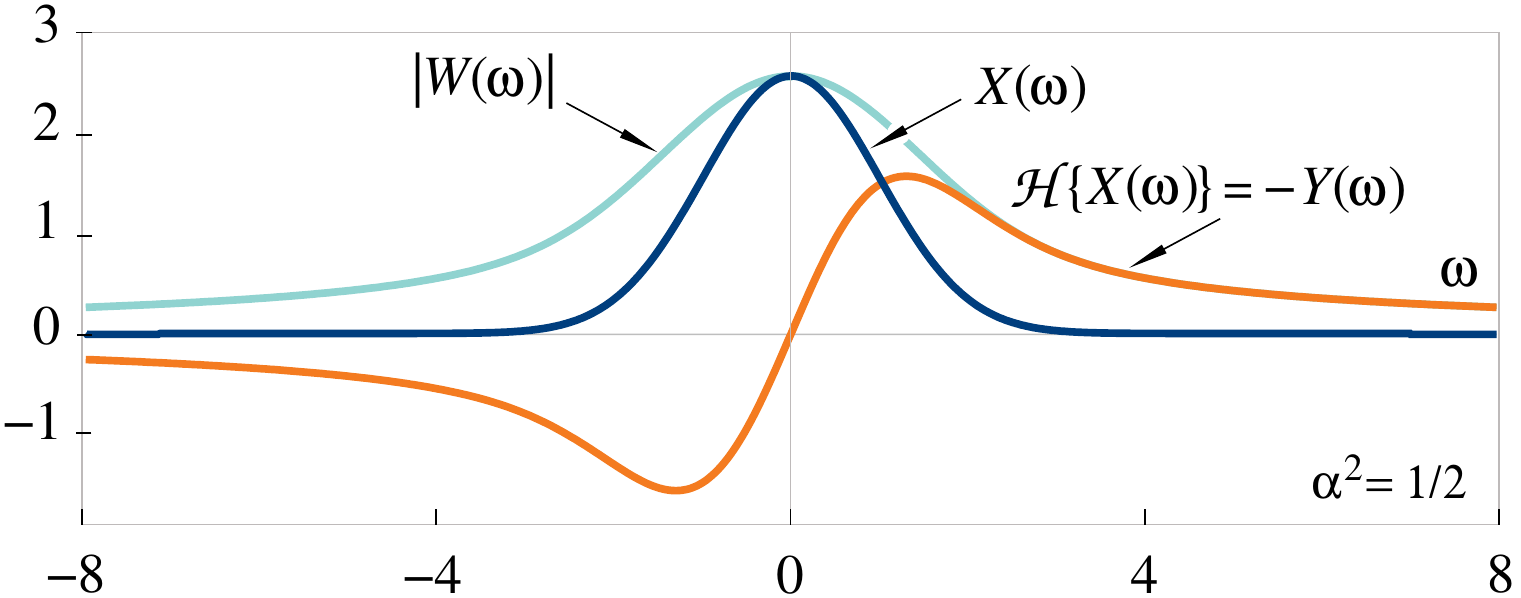}%{ARIGA.pdf}
	\vspace{-0.3cm}
	\caption{Real part, $X(\omega)$, its Hilbert transform, $\mathcal{H}\{X(\omega)\} = -Y(\omega)$, and spectral envelope, $|W(\omega)|$, of the complex Fourier spectrum of a causal Gaussian pulse.}
\end{figure}

\vspace{0.1cm}
\noindent {\emph{Example 1:}} Consider a causal form of a Gaussian pulse
\begin{equation}
w(t)  \,=\, \exp(-\alpha^2\mspace{1mu} t^2)[1+ \mathrm{sgn}(t)],\quad \alpha > 0.
\end{equation}
The two components, $X(\omega)$ and $Y(\omega)$, of its complex Fourier spectrum, $W(\omega)$, are given by
\begin{eqnarray}
	X(\omega)\mspace{-9mu}  &=&\mspace{-7mu} (\sqrt{\pi}/a)\mspace{1mu} \exp\mspace{1mu}[-\omega^2/(4 a^2)]\\ \nonumber
	Y(\omega)\mspace{-9mu}  
	&=&\mspace{-7mu} j(\sqrt{\pi}/a)\mspace{1mu} \exp
	\mspace{1mu}[\mspace{1mu}-\omega^2/(4 a^2)] \mspace{2mu}
	\mathrm{erf}\mspace{1mu}[\mspace{2mu}j\omega/(2a)] 	\\ 	    
	&=&\mspace{-7mu} -(2/a) \mspace{1mu}	    \mathbb{D}\mspace{-2mu}\left[\mspace{1mu}\omega/(2a) \right]
\end{eqnarray}
where $\mathrm{erf}(\cdot)$ is the error function [7, (7.2.1)], 
\begin{equation}
	\mathrm{erf}(\zeta)  \,\triangleq\,
	(2/\sqrt{\pi})\mspace{-3mu} \int_0^\zeta \mspace{-5mu}
	\exp(-\rho^2) \,d\rho
\end{equation}
and $\mathbb{D}\{\cdot\}$ is the Dawson integral [7, (7.2.5)],
\begin{equation}
	{\mathbb{D}}\!\left( \zeta \right) \,\,\triangleq \,\,
	\exp(-\zeta^2)\!\!  \int_0^\zeta  \!  \exp(\mspace{1mu}\rho^2) \,d\rho. \nonumber
\end{equation}
Since [8], $\mathcal{H}\{X(\omega)\} = 
	(2/a) \mspace{1mu}	    \mathbb{D}\mspace{-2mu}\left[\mspace{1mu}\omega/(2a) \right]$,
 the relationship (14) is satisfied.

Components of the complex Fourier spectrum $W(\omega)$ and the resulting spectral envelope,  
$$
|W(\omega)| \,=\,\sqrt{X^2(\omega)+Y^2(\omega)}
$$
are shown in Fig.\,4. %As seen, the spectral component $Y(\omega)$ occupies a wider frequency band than that occupied %by $X(\omega)$.

\begin{figure}[] %[b]
	\centering
	\includegraphics[width=7.5cm]{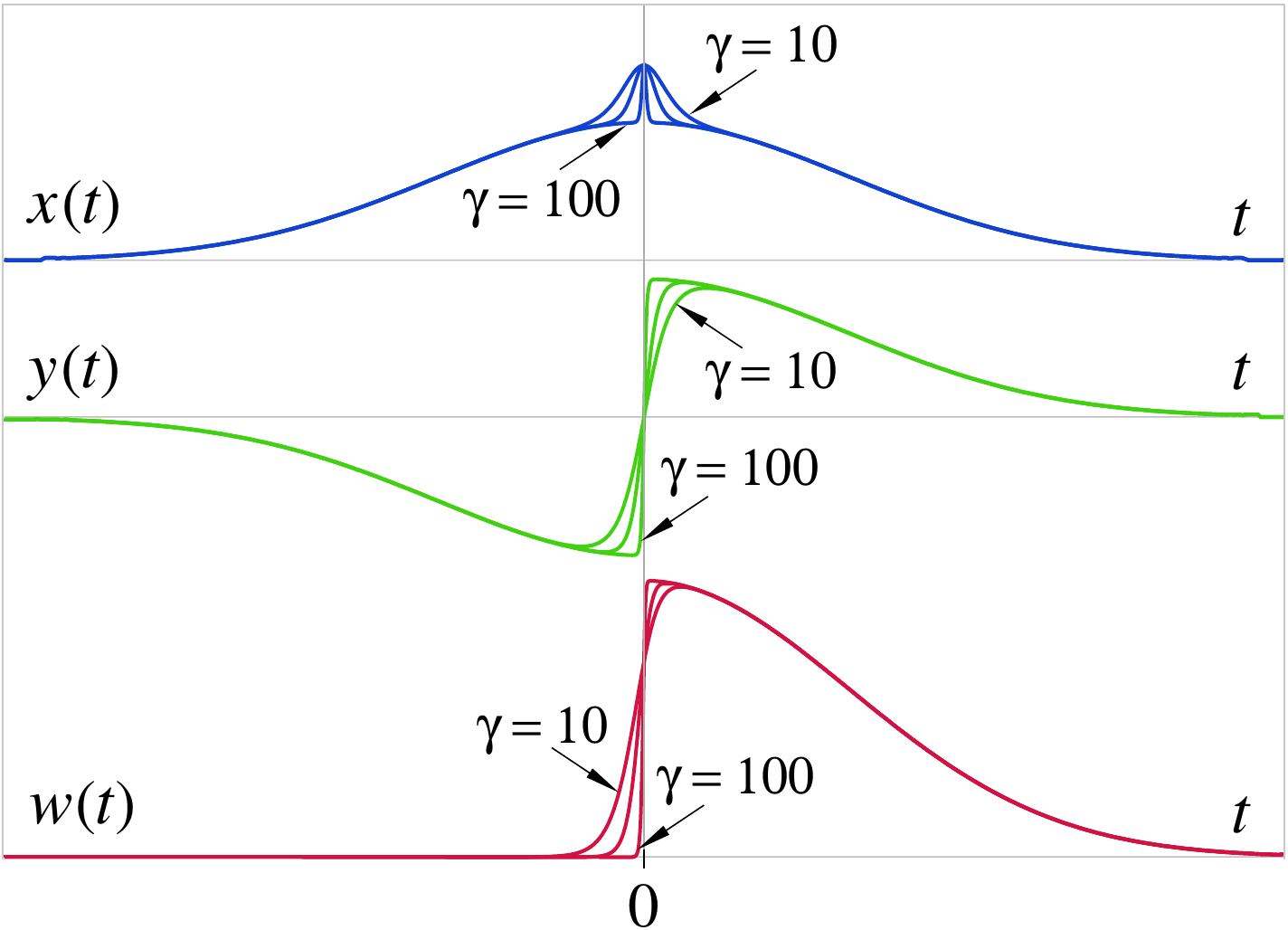}
	\vspace{-0.3cm}
	\caption{Symmetric pulse waveform, $x(t)$, its  antisymmetric replica, $y(t)$, and the sum, $w(t)$, for $\gamma = 10, 20, 100$.}
\end{figure}

\subsection{Unimodal and Bimodal Pulses} 
In the case of minimal pulse compression, the phase modulating function (9) assumes the form
\begin{eqnarray}
	\psi(t) = \mspace{1mu}  (\pi/4)\mspace{1mu}\mathrm{sgn}(t).
\end{eqnarray}
Since the instantaneous change of phase from $-\pi/4$ to $+\pi/4$ cannot be achieved in a physical system, in a more practical model, the signum function appearing in (19) may be replaced by some suitably chosen sigmoid function.   

For example, the mathematical model (19) of a phase modulating function may be modified as follows
\begin{eqnarray}
	\psi(t) = \mspace{1mu}  (\pi/4) \tanh(\gamma \mspace{2mu} t)
\end{eqnarray}
where $\gamma$ is a shape parameter; when $\gamma$ approaches infinity, 
$\tanh(\gamma\mspace{1mu} t)$ tends to $\mathrm{sgn}(t)$. Consequently, the two pulse components are given by
\begin{eqnarray}
	x(t)\!\! &\!\!=\!\!& \!\mu(t) \cos\mspace{1mu}[\pi \tanh(\gamma \mspace{2mu} t)/4]
	\nonumber \\
	y(t)\!\! &\!\!=\!\!& \!  
	\mu(t) \sin\mspace{1mu}[\pi\tanh(\gamma \mspace{2mu} t)/4]
\end{eqnarray}
and their sum, $w(t)$, will  no longer be a causal waveform. 

Fig.\,5 shows two components, $x(t)$ and $y(t)$, of a Gaussian pulse for three different values of shape parameter $\gamma$. For large values of $\gamma$, the sum $w(t)$ can be regarded as a good approximation of a causal pulse.

For illustration purposes, it is convenient to represent an RF pulse by a time-varying phasor with two orthogonal components, $x(t)\cos(\omega_0\mspace{1mu}t)$ and $y(t) \sin(\omega_0\mspace{1mu}t)$.

Fig.\,6 shows a helix trajectory traced by the tip of the phasor; at $t=0$, the trajectory  changes its angular direction from clockwise to anticlockwise. It should be noted that the shape of the envelope, $\mu(t) = |z(t)|$, remains unchanged, irrespective of the value of $\gamma$.

\begin{figure}[] %[b]
	\centering
	\includegraphics[width=7.3cm]{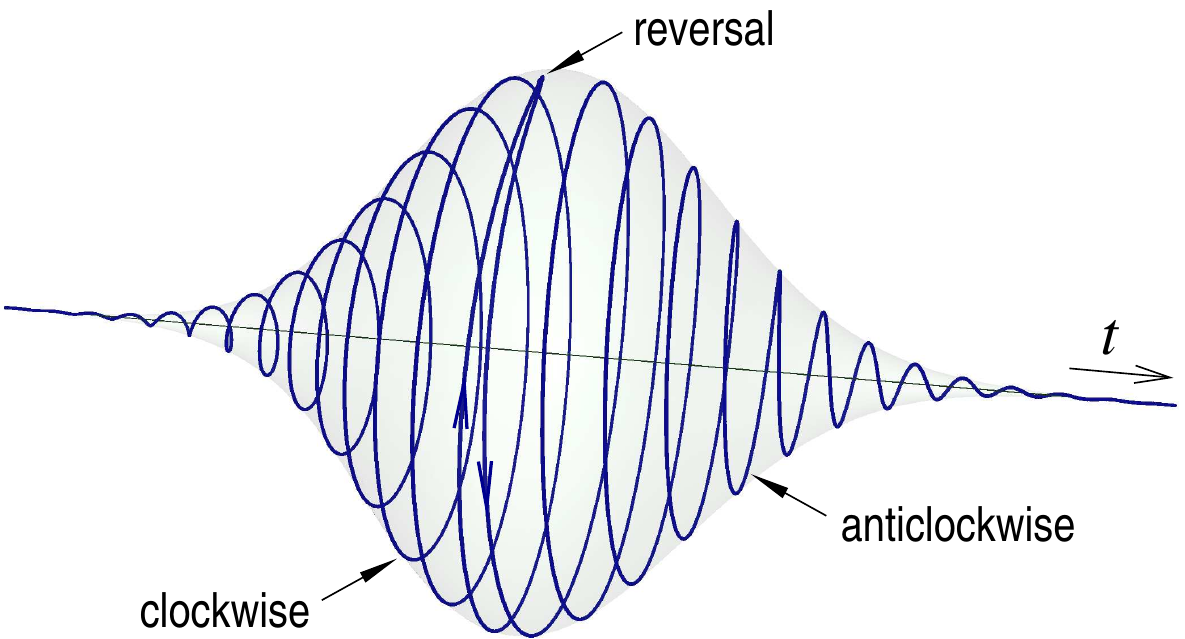}
	\vspace{-0.1cm}
	\caption{Trajectories traced by the tip of a phasor representing a minimally compressible Gaussian RF pulse.}
\end{figure}

\subsubsection{Pulses with Bimodal Envelope} 
A minimally compressive pulse can also be created by using an antisymmetric component, $y(t)$, as a primary waveform, to form the other (symmetric) component as
\begin{equation}
	x(t) = y(t) \mspace{2mu}\mathrm{sgn}(t)\mspace{1mu}.
\end{equation}  
In such a case, the waveform $y(t)$ is assumed to be physically realizable, and the resulting pulse envelope will no longer have a unimodal shape.

One way to create a mathematical model of an antisymmetric waveform with controllable steepness of the slope at $t=0$ is to multiply a symmetric waveform, such as one of those shown in Fig.\,1, by a sigmoid function. Another approach may be based on using a physically realizable antisymmetric waveform. 

Irrespective of the method, such created in-phase and quadrature components, $x(t)$ and  $y(t)$, are used to form an RF pulse,
\begin{equation*}
	s(t) = 	x(t)\cos(\omega_0\mspace{1mu}t) \mp
	y(t) \sin(\omega_0\mspace{1mu}t).
\end{equation*}

\begin{figure}[t] %[b]
	\centering
	\includegraphics[width=7.5cm]{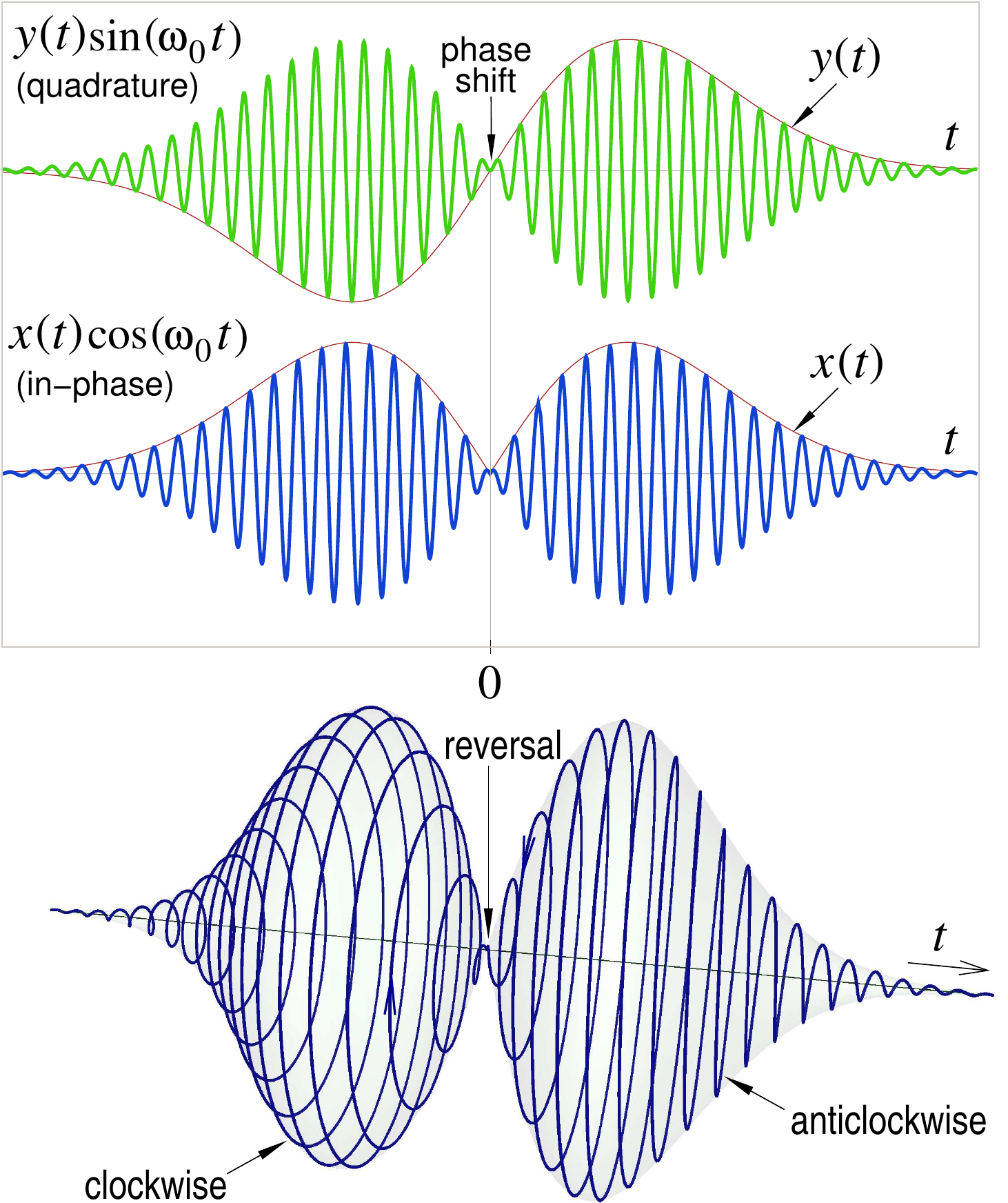}
	\vspace{-0.2cm}
	\caption{Minimally compressible Hermite-Gaussian pulse: quadrature and in-phase components (top), trajectory traced by the phasor's tip (bottom).}
\end{figure}

\subsubsection*{Hermite-Gaussian Pulse}
As an example of a physically realizable antisymmetric waveform, consider 
a Hermite-Gaussian bipolar pulse [9],
\begin{equation}
	y(t) = t \mspace{1mu}\exp\mspace{1mu}(-\lambda^2 t^2/2), \quad \lambda>0.
\end{equation} 
and a corresponding symmetric non-negative pulse,
\begin{equation}
	x(t) = |t| \mspace{1mu}\exp\mspace{1mu}(-\lambda^2t^2/2).
\end{equation} 
Accordingly, the Hermite-Gaussian causal pulse assumes the form
\begin{equation}
	w(t)  \,=\, |t| \mspace{1mu}\exp\mspace{1mu}(-\lambda^2t^2/2)[1+ \mathrm{sgn}(t)].
\end{equation}

Fig.\,7 (top) shows quadrature and in-phase components of a Hermite-Gaussian pulse, and also waveforms, given by (23) and (24), respectively. 

Fig.\,7 (bottom) depicts a helix trajectory traced by the tip of a corresponding phasor; at $t=0$, the trajectory changes its angular direction from clockwise to anticlockwise.

Two components, $X(\omega)$ and $Y(\omega)$, of the complex Fourier spectrum, $W(\omega)$, of $w(t)$ are given by
%{\footnote {\,The real part, $X(\omega)$, can also be expressed in terms of the Dawson integral.}}
\begin{eqnarray}
	X(\omega)\mspace{-10mu}  &=&\mspace{-8mu} 
	(\sqrt{2\pi}/\lambda^3) \mspace{1mu}\omega \exp\mspace{1mu}[\mspace{1mu}-\mspace{1mu}\omega^2\mspace{-2mu}
	/(2\mspace{1mu} \lambda^2)] \mspace{1mu}j\mspace{1mu}
	\mathrm{erf}\mspace{1mu}[\mspace{2mu}j\mspace{1mu}\omega/(\sqrt{2}\mspace{1mu}\lambda)] 
	\nonumber		\\ 	    &{}&\mspace{-8mu} +\, 2/\lambda^2
	\\ 
	Y(\omega)\mspace{-10mu}  &=&\mspace{-8mu} -(\sqrt{2\pi}/\lambda^3)\mspace{2mu} \omega  
	\exp\mspace{1mu}[\mspace{1mu}-\mspace{1mu}\omega^2\mspace{-2mu}
	/(2 \mspace{1mu}\lambda^2)].
	\end{eqnarray}
It can be shown that also in this case, $Y(\omega)=-\mathcal{H}\{X(\omega)\}$, so that the relationship (14) is satisfied. 

Fig.\,8 shows two components of the complex Fourier spectrum, $W(\omega)$, and the resulting  spectral envelope, $|W(\omega)|$. 

A minimally compressible Hermite-Gaussian pulse can be viewed as being created by abutting two pulses having mirror-image envelopes and quadrature components of opposite signs.

\begin{figure}[t] %[b]
	\centering
	\includegraphics[width=8.6cm]{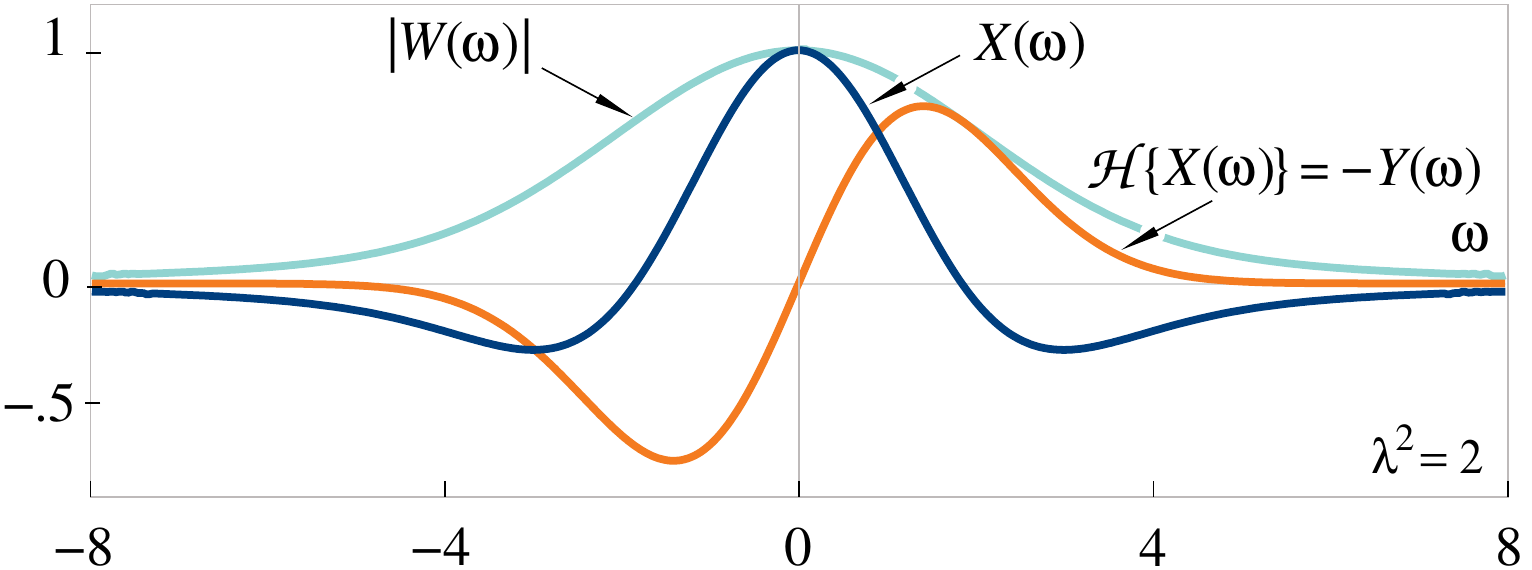}%{ARIGA.pdf}
	\vspace{-0.3cm}
	\caption{Real part, $X(\omega)$, its Hilbert transform, $\mathcal{H}\{X(\omega)\} = -Y(\omega)$, and spectral envelope, $|W(\omega)|$, of the complex Fourier spectrum of a causal Hermite-Gaussian pulse.}
\end{figure}

\subsection{Autocorrelation of a Complex Envelope} 
When a complex signal, $z(t)= x(t)+jy(t)$, of known form is to be detected in additive wideband noise, the resulting combination of signal and noise is usually applied to a complex matched filter or a correlator. A signal component observed at the output of either device assumes the form of the autocorrelation function $R_{zz}(\tau)$, 
\begin{eqnarray}
	R_{zz}(\tau) &\!\!\triangleq\!\!& \! 
	\int_{-\infty}^{\infty}\!\! {z}^*(t)\mspace{1mu} z(t+\tau) \mspace{2mu} dt \nonumber \\
	&\!\!=\!\!\!& \!
	\int_{-\infty}^{\infty}\!\!
	[x(t) - j y(t)]\mspace{1mu}[x(t+\tau) + j y(t+\tau)] \mspace{2mu} dt  \qquad 
\end{eqnarray}
where $*$ denotes the complex conjugate.

Accordingly, the real and imaginary parts of the autocorrelation function $R_{zz}(\tau)$ are given by$\mspace{1mu}${\footnote {\,For the conjugate, $z^*(t)=x(t)\!-\!jy(t)$, 
$\mathcal{R}e\{R_{z^{\mspace{-2mu}*}z^{\mspace{-2mu}*}}(\tau)\}=R_\Sigma(\tau)$ and 
$\mathcal{I}m\{R_{z^{\mspace{-2mu}*}z^{\mspace{-2mu}*}}(\tau)\}=-R_\Delta(\tau)$; 
hence, $R_{z^{\mspace{-2mu}*}z^{\mspace{-2mu}*}}(\tau)= R^*_{zz}(\tau)$
}} 
\begin{eqnarray}
	R_\Sigma(\tau)\!\! &\!\triangleq\!\!& 
	\mathcal{R}e\{R_{zz}(\tau)\}\,=\,
	R_{xx}(\tau) + R_{yy}(\tau)  \\
	R_\Delta(\tau)\!\! &\!\triangleq\!\!&   
	\mathcal{I}m\{R_{zz}(\tau)\}\,=\,
	R_{xy}(\tau) - R_{yx}(\tau). \quad
\end{eqnarray}

Since, the autocorrelation function, $R_{xx}(\tau)$, of a pulse $x(t)$ is defined by
\begin{equation}
	R_{xx}(\tau)\, \triangleq \, 
	\int_{-\infty}^{\infty}\!\!
	x(t)\mspace{1mu} x(t+\tau) \, dt
\end{equation}
the autocorrelation function, $R_{yy}(\tau)$, of the antisymmetric replica $y(t)$ of $x(t)$ can be determined from
\begin{eqnarray}
	R_{yy}(\tau)\!\! &\!\!\triangleq\!\!& \! \!
	\int_{-\infty}^{\infty}\!\!
	y(t)\mspace{1mu} y(t+\tau) \, dt \nonumber \\
	&=& \!\!
	\int_{-\infty}^{\infty}\!\!
	x(t)\mspace{1mu} x(t+\tau)\mspace{1mu}
	\mathrm{sgn}(t)\mspace{1mu} \mathrm{sgn}(t+\tau) \, dt. \quad
\end{eqnarray}
While the autocorrelation function $R_{xx}(\tau)$ of a (positive) pulse $x(t)$ may only assume non-negative values, no such restriction applies to the autocorrelation function $R_{yy}(\tau)$ of the antisymmetric pulse $y(t)$. 
\begin{figure}[t] %[b]
	\centering
	\includegraphics[width=8.5cm]{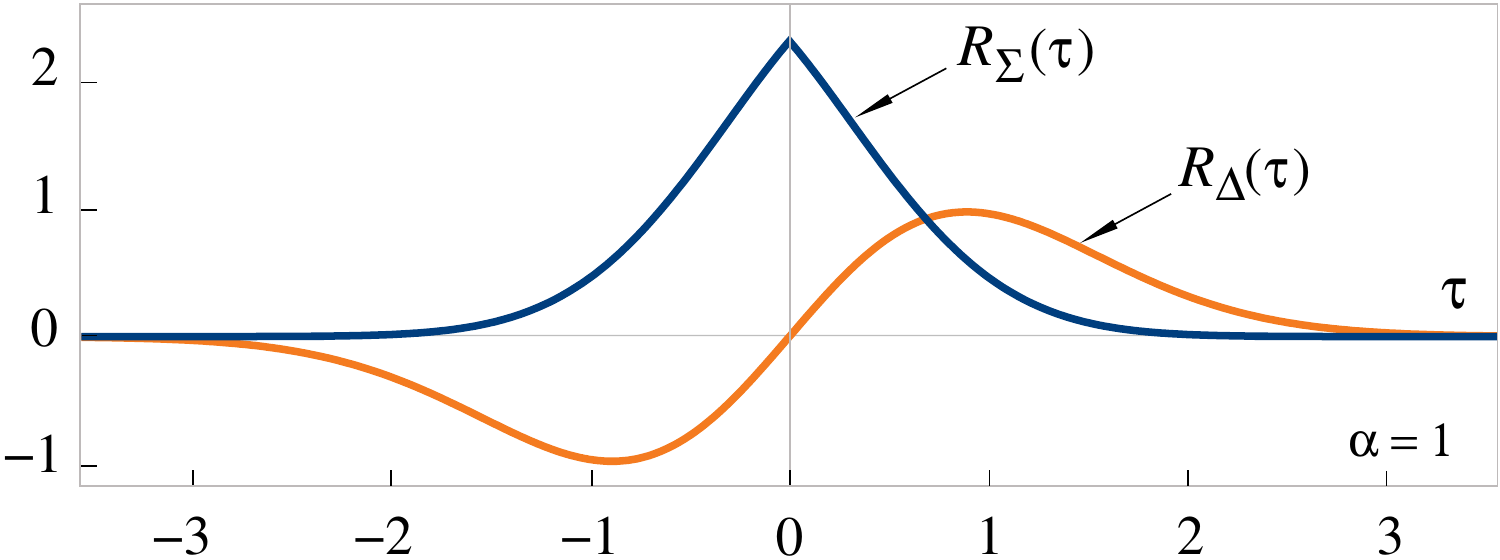}%{ARIGA.pdf}
	\vspace{-0.3cm}
	\caption{Real and imaginary parts, $R_\Sigma(\tau)$ and $R_\Delta(\tau)$, of the autocorrelation function of a minimally compressible Gaussian pulse.}
\end{figure}
It can be shown (see Appendix) that 
\begin{eqnarray}
	R_\Sigma(\tau)\! &\!\!=\!\!& 4 \mspace{-4mu}
	\int_{0}^{\infty} \mspace{-5mu}
	x(t)\mspace{1mu} x(t+|\tau|) \, dt \nonumber\\
	\!&\!\!=\!\!& 4 \mspace{-4mu}
	\int_{0}^{\infty} \mspace{-5mu}
	y(t)\mspace{1mu} y(t+|\tau|) \, dt \nonumber\\
	\!&\!\!=\!\!& 
	\int_{0}^{\infty} \mspace{-5mu}
	w(t)\mspace{1mu} w(t+|\tau|) \, dt .\quad
\end{eqnarray} 
and
\begin{eqnarray}
	R_\Delta(\tau)\! &\!\!=\!\!&   2\mspace{3mu} \mathrm{sgn}(\tau) \mspace{-4mu}
	\int_{-|\tau|}^{\mspace{1mu}0} \mspace{-5mu}
	x(t)\mspace{1mu} x(t+|\tau|) \, dt \nonumber \\
	\!&\!\!=\!\!&   -2\mspace{3mu} \mathrm{sgn}(\tau) \mspace{-4mu}
	\int_{-|\tau|}^{\mspace{1mu}0} \mspace{-5mu}
	y(t)\mspace{1mu} y(t+|\tau|) \, dt	.\,\,
\end{eqnarray} 

The function $R_\Sigma(\tau)$ is an {\emph {even}} function of time delay $\tau$ and it may only assume non-negative values. Therefore, the function $R_\Sigma(\tau)$ may be used for localizing a pulse in time. On the other hand, the imaginary part, $R_\Delta(\tau)$, of the complex autocorrelation function $R_{zz}(\tau)$ is an {\emph {odd}} function of time delay $\tau$, and $R_\Delta(0)=0\mspace{1mu}$; hence, the function $R_\Delta(\tau)$ may be exploited for delay tracking in a closed-loop system. 

\vspace{0.1cm}
\noindent {\emph{Example 2:}} Consider a complex representation,
\begin{equation}
	z(t)  \,=\, \exp(-\alpha^2\mspace{1mu} t^2)[1+ j\mspace{2mu}\mathrm{sgn}(t)],
	\quad \alpha > 0.
\end{equation}
of a minimally compressible Gaussian pulse (15).

The real and imaginary parts, $R_\Sigma(\tau)$ and $R_\Delta(\tau)$, of the autocorrelation function, $R_{zz}(\tau)$, are given by
\begin{eqnarray}
	R_\Sigma(\tau) \!\!&\!\!=\!\!&\!  \sqrt{2 \pi} \exp(-\alpha^2 \tau^2/2) 
	\mspace{2mu} \mathrm{erfc}(\alpha\mspace{1mu} |\tau|/\sqrt{2})/\alpha    \\
	R_\Delta(\tau) \!\!&\!\!=\!\!&
		\! \mathrm{sgn}(\tau)\sqrt{2 \pi} \exp(-\alpha^2 \tau^2/2) 
	\mspace{2mu} \mathrm{erf}(\alpha\mspace{1mu} |\tau|/\sqrt{2})/\alpha 
	\quad \, \nonumber   \\
	&\!\!=\!\!&	\! \sqrt{2 \pi} \exp(-\alpha^2 \tau^2/2) 
	\mspace{2mu} \mathrm{erf}(\alpha\mspace{1mu} \tau/\sqrt{2})/\alpha \nonumber 
\end{eqnarray}
where $\mathrm{erfc}(\cdot)$ is the complementary error function [7, (7.2.2)]
\begin{equation}
	\mathrm{erfc}(\zeta)  \,\triangleq\,
	(2/\sqrt{\pi})\mspace{-3mu} \int_\zeta^\infty \mspace{-5mu}
	\exp(-\rho^2) \,d\rho
\end{equation}
and the error function $\mathrm{erf}(\cdot)$ is given by (18). 

The value $R_\Sigma(0)$ is equal to $\sqrt{2 \pi}/\alpha$, i.e. the total energy of the pulse (35). It should be noted that the function $R_\Sigma(\tau)$ has a cusp at $\tau=0$.

The components of the complex autocorrelation function $R_{zz}(\tau)$, for $\alpha=1$, are shown in Fig.\,9. 

\vspace{0.15cm}
\subsubsection{Frequency-Domain Approach}
From the Wiener-Khintchin theorem, it follows that
\begin{eqnarray}
	R_{xx}(\tau)\!\! &\!\!=\!\!& \! \mathcal{F}^{-1}\{|X(\omega)|^2\} 
	\nonumber \\
	R_{yy}(\tau)\!\! &\!\!=\!\!& \! \mathcal{F}^{-1}\{|Y(\omega)|^2\}
	= \mathcal{F}^{-1}\{[\mathcal{H}\{X(\omega)\}]^2\} .\quad
	\end{eqnarray}
Consequently, the real and imaginary parts of the autocorrelation function $R_{zz}(\tau)$ can be determined from the Fourier spectra, $X(\omega)$ and $Y(\omega)$, of the respective components. Therefore,
\begin{eqnarray}
	R_\Sigma(\tau) \! &\!\!=\!\!& \!
	\mathcal{F}^{-1}\{|X(\omega)|^2+|Y(\omega)|^2\} \nonumber \\
	\! &\!\!=\!\!& \!
	\mathcal{F}^{-1}\{|W(\omega)|^2\} 
\end{eqnarray}
${}$
\vspace{-0.7cm}

\noindent
and
\begin{equation}
	R_\Delta(\tau)\,\,=\,\, \mathrm{sgn}(\tau)\mspace{3mu}
	\mathcal{F}^{-1}\{|X(\omega)|^2-|Y(\omega)|^2\}.\!\!
\end{equation}

\vspace{0.05cm}
\noindent {\emph{Example 3:}} Consider a complex representation,
\begin{equation}
	z(t)  = \exp(-\beta\mspace{1mu}|t|)[1+ j\mspace{2mu}\mathrm{sgn}(t)],\quad \beta>0.
\end{equation}
of a minimally compressible Laplacian pulse. 

The Fourier spectra of real and imaginary parts of $z(t)$, are given by
\begin{eqnarray}
	X(\omega) \!\!&\!\!=\!\!&\!\!  2\mspace{2mu} \beta/(\omega^2+\beta^2) \nonumber \\
	Y(\omega) \!\!&\!\!=\!\!&\!\! 2\mspace{2mu} \omega/(\omega^2+\beta^2). 
\end{eqnarray}
Therefore, the real and imaginary parts, $R_\Sigma(\tau)$ and $R_\Delta(\tau)$, of the autocorrelation function, $R_{zz}(\tau)$, determined from (39) and (40), can be expressed as
\begin{eqnarray}
R_\Sigma(\tau) \! &\!\!=\!\!& \!
4\, \mathcal{F}^{-1}\{1/(\omega^2+\beta^2)\} 
\,=\, 2\mspace{1mu} \exp(-\beta |\tau|)/\beta   \\	
R_\Delta(\tau) \! &\!\!=\!\!& \!	
	4\,\mathrm{sgn}(\tau)\mspace{3mu}\mathcal{F}^{-1}\{-(\omega^2-\beta^2)/
	(\omega^2+\beta^2)^2\} \nonumber \\
	\! &\!\!=\!\!& \!2\,\mathrm{sgn}(\tau)\mspace{2mu}|\tau| \exp(-\beta |\tau|)
	\,=\, 2\mspace{1mu}\tau \exp(-\beta |\tau|).\nonumber \,\qquad
\end{eqnarray}
The even function $R_\Sigma(\tau)$ has a cusp at $\tau=0$, and its peak value $R_\Sigma(0)$ is equal to $2/\beta$, i.e. the total energy of the pulse (41); the function $R_\Delta(\tau)$ is an odd function of delay $\tau$.
\begin{figure}[t] %[b]
	\centering
	\includegraphics[width=8.6cm]{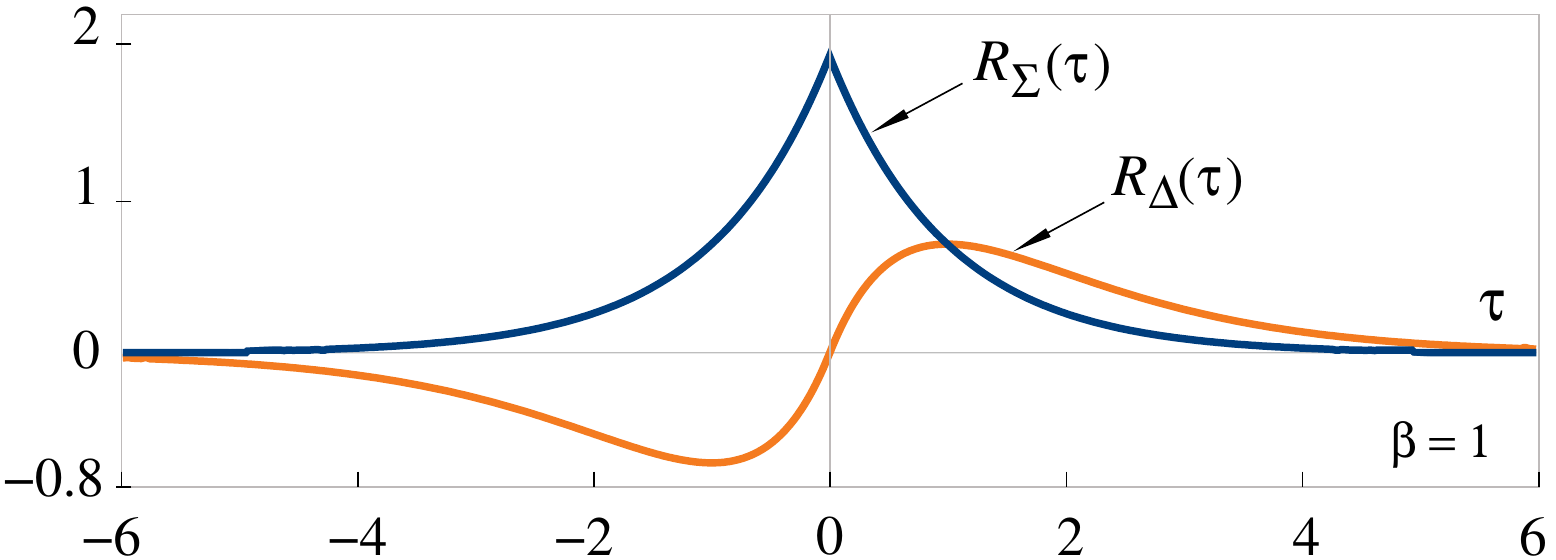}%{ARIGA.pdf}
	\vspace{-0.3cm}
	\caption{Real and imaginary parts, $R_\Sigma(\tau)$ and $R_\Delta(\tau)$, of the autocorrelation function of a minimally compressible Laplacian pulse.}
\end{figure}

The real and imaginary parts, $R_\Sigma(\tau)$ and $R_\Delta(\tau)$, of the complex autocorrelation function $R_{zz}(\tau)$, for $\beta=1$, are shown in Fig.\,10.

\section{Resolution Gain and Correlation Processing} 
The problem of minimal pulse compression is somewhat related to exploiting a pair of   complementary (Golay) binary sequences in radar [2]. Addition of respective autocorrelation functions of two corresponding complementary sequences results in a function having a single peak at zero delay and zero-level sidelobes at all other delays.

Fig.\,11\,a shows a pair of shortest complementary Golay sequences, 
$(+1,+1)$ and $(-1,+1)$, their respective autocorrelation sequences, $(+1,+2,+1)$ and $(-1,+2,-1)$, and the resulting sum, $(0,+4,0)$, of the autocorrelation sequences.

Corresponding binary waveforms, their autocorrelation functions, and the resulting sum of these functions are shown in Fig.\,11\,b. The two binary waveforms may be regarded as a special case of a symmetric pulse and its antisymmetric replica (see Fig.\,2). 

\subsection{Resolution Gain} 
In radar systems, a pulse compression gain $G_C$ is usually defined as the ratio, 
\begin{equation}
G_C \, \triangleq\,  T_O/T_C	
\end{equation}
where $T_O$ and $T_C$ are the durations of matched-filter responses to, respectively, an RF pulse and the same pulse modified by some kind of intra-pulse modulation, such as phase or frequency modulation. 

The duration of each response is assumed to be equal to its FWHH (full-width half-height) value. For example, in the case depicted in Fig.\,11\,b, $G_C=2$.

In the following, the autocorrelation functions, $R_{xx}(\tau)$, $R_{yy}(\tau)$, their sum, $R_\Sigma(\tau)$, and the resulting compression gain, $G_C$, will be determined for the three different pulse shapes shown in Fig. 1, and also for a bimodal Hermite-Gaussian pulse, discussed in Section III-B.

\vspace{0.1cm}
\subsubsection{Laplacian Pulse}
Consider a Laplacian pulse, 
\begin{equation}
	 x(t) \,= \,\exp(-\beta\mspace{1mu}|t|), \quad \beta > 0.
\end{equation}
When its autocorrelation function,
\begin{equation}
	R_{xx}(\tau) \,= \,(1+\beta \mspace{1mu}|\tau|)\exp(-\beta\mspace{1mu}|\tau|)/\beta
\end{equation}
is added to that of a corresponding antisymmetric pulse, 
\begin{equation}
	R_{yy}(\tau) \,= \,(1-\beta \mspace{1mu}|\tau|)\exp(-\beta\mspace{1mu}|\tau|)/\beta
\end{equation}
the resulting function, $R_\Sigma(\tau)$, given by (43), is obtained.

Normalised forms, $R^o_{xx}(\tau)$, $R^o_{yy}(\tau)$ and $R^o_\Sigma(\tau)$, of the respective functions are shown in Fig.\,12\,a. 
In this case, the compression gain, $G_C \approx 2.42$.

\vspace{0.05cm}
\subsubsection{Gaussian Pulse}
Consider a Gaussian pulse,
\begin{equation}
	x(t) \,= \,\exp(-\alpha^2\mspace{1mu} t^2), \quad \alpha > 0 .
\end{equation}
Its autocorrelation function, $R_{xx}(\tau)$, is of the form
\begin{equation}
	R_{xx}(\tau) \,= \,\sqrt{\pi/2} \exp(-\alpha^2 \tau^2/2)/\alpha
\end{equation}
and the autocorrelation function, $R_{yy}(\tau)$, of the corresponding antisymmetric pulse, $y(t)$, can be expressed as
\begin{eqnarray}
	R_{yy}(\tau) \!\!&\!\!=\!\!&\!\!  
	\sqrt{\pi/2}\mspace{1mu} \exp(-\alpha^2 \tau^2/2)\mspace{-3mu}
	\left[2\mspace{1mu}\mathrm{erfc}(\alpha\mspace{1mu} |\tau|/\sqrt{2})\!-\!1\right]\mspace{-1mu}
	\big/\alpha 	\nonumber \\
	&\!\!=\!\!&\!\! 
	\sqrt{\pi/2}\mspace{1mu} \exp(-\alpha^2 \tau^2/2)\mspace{-3mu}
	\left[1\!-\!2\mspace{1mu}\mathrm{erf}(\alpha\mspace{1mu} |\tau|/\sqrt{2})\right]\mspace{-1mu}
	\big/\alpha. \nonumber  
\end{eqnarray}
Consequently, their sum, $R_\Sigma(\tau)$, is given by (36), and the resulting compression gain, $G_C \approx 2.1$.

Normalised forms, $R^o_{xx}(\tau)$, $R^o_{yy}(\tau)$ and $R^o_\Sigma(\tau)$, of the respective functions are shown in Fig.\,12\,b. 

\begin{figure}[] %[b]
	\centering
	\includegraphics[width=8.8cm]{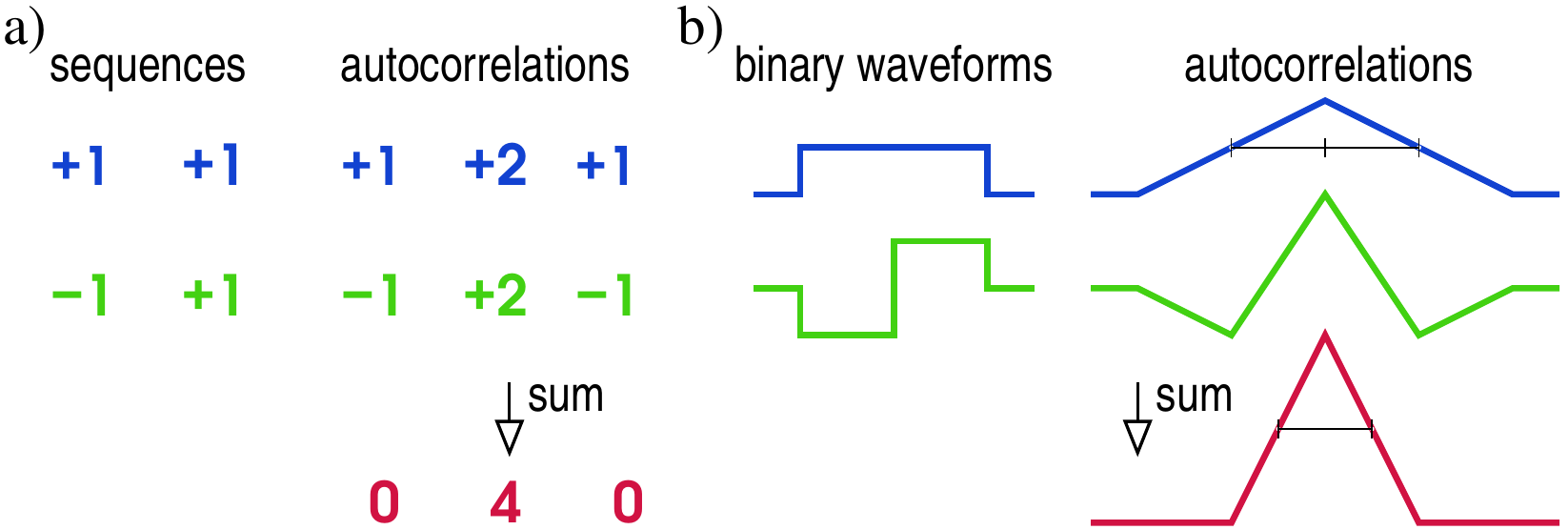}%{ARIGA.pdf}
	\vspace{-0.25cm}
	\caption{A pair of shortest complementary sequences, corresponding binary waveforms and resulting autocorrelations.}
\end{figure}

\vspace{0.05cm}
\subsubsection{'Soft' Rectangular Pulse}
A 'soft' rectangular pulse (2) is of the form, 

$\qquad x(t) \,=\, 
	\{\tanh\mspace{1mu}[\gamma\mspace{1mu} (t\mspace{-1mu}+\mspace{-1mu}\kappa)]
	-\tanh\mspace{1mu}[\gamma\mspace{1mu}(t\mspace{-1mu}-\mspace{-1mu}\kappa)]
	\mspace{1mu}\}\mspace{-1mu} /2 $\\
where $-\kappa$ and $+\kappa$ are nominal positions of its edges, and the shape parameter $\gamma$ characterises their steepness.

When the value of $\gamma$ approaches infinity, a 'soft' rectangular pulse becomes a rectangular pulse, 

$\qquad x(t)=
[\mspace{1mu}\mathrm{sgn}\mspace{1mu} (t\mspace{-1mu}+\mspace{-1mu}\kappa) - \mathrm{sgn}\mspace{1mu}(t\mspace{-1mu}-\mspace{-1mu}\kappa)]/2\mspace{1mu}$,\\
 with a triangular autocorrelation function,
\begin{equation}
	R_{xx}(\tau) \,=\, 
	{\Lambda}\mspace{1mu} (\tau; 2\mspace{1mu}\kappa) 
\end{equation}

\vspace{-0.2cm}
\noindent where 

\vspace{-0.4cm}
\begin{equation}
	{\Lambda}\mspace{1mu} (\tau; \rho) 
	\,\, \triangleq \,\,
	\left\{
	\begin{array}{l@{\,\,\,\,\,\,\,\,\,}l} 
		{\mspace{-4mu}   1- {|\tau|}/{\rho}  ,} 
		&{|\tau|\leq \rho}\\
		{\mspace{-4mu}0,} & {|\tau| > \rho}\, .
	\end{array} \right .
\end{equation}

The autocorrelation function, $R_{yy}(\tau)$, of the corresponding antisymmetric pulse, 
$y(t)=x(t)\mspace{1mu} \mathrm{sgn}(t)$, assumes the form
\begin{equation}
	R_{yy}(\tau) \,= \, 2\mspace{1mu}{\Lambda}\mspace{1mu} (\tau; \kappa) - 
	{\Lambda}\mspace{1mu} (\tau; 2\mspace{1mu}\kappa) .
\end{equation}
Therefore,
\begin{equation}
	R_\Sigma(\tau) \,= \,2\mspace{1mu}{\Lambda}\mspace{1mu} (\tau; \kappa) 
\end{equation}
and the resulting compression gain, $G_C = 2$, represents the smallest achievable value.  

The shapes of the three piecewise linear functions (50), (52) and (53) are depicted in Fig.\,11\,b.

\begin{figure}[] %[b]
	\centering
	\includegraphics[width=8.7cm]{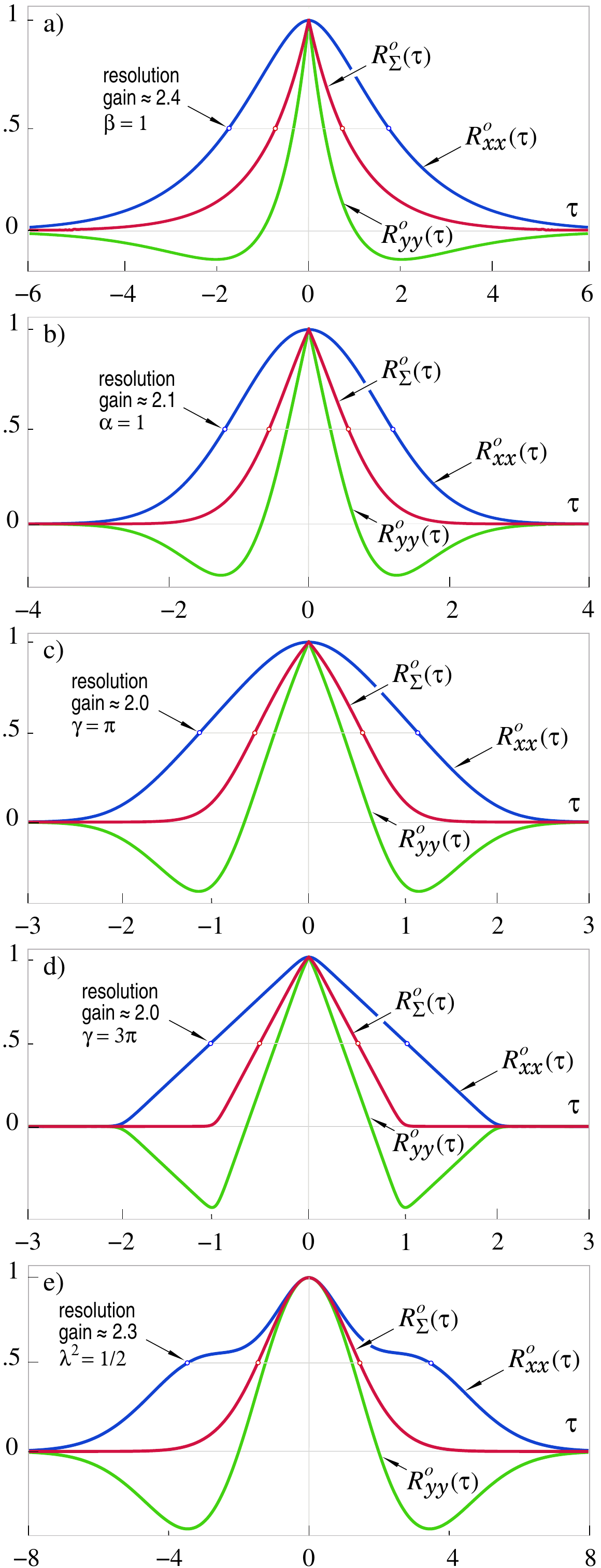}
	\vspace{-0.2cm}
	\caption{Normalised autocorrelation functions: $R^o_{xx}(\tau)$, $R^o_{yy}(\tau)$ and their normalised sum $R^o_\Sigma(\tau)$ of a minimally compressible pulse with a specified shape: Laplacian (a); Gaussian (b); 'soft' rectangular, (c) and (d); Hermite-Gaussian (e).}
\end{figure}

In a more general case, when $\gamma \neq \infty$, each of the three autocorrelation functions, $R_{xx}(\tau)$, $R_{yy}(\tau)$ and $R_\Sigma(\tau)$, can be expressed in a closed form. However, the expressions derived with the use of {\tt {Maxima}}, a computer algebra system,  are very long and thus not well suited to the publication format.

When the shape parameter value, $\gamma$, is gradually decreasing from infinity, the shapes of the respective autocorrelation functions, $R_{xx}(\tau)$, $R_{yy}(\tau)$ and $R_\Sigma(\tau)$, 
are starting to depart more and more noticeably from those of piecewise linear functions.

 Normalised forms, $R^o_{xx}(\tau)$, $R^o_{yy}(\tau)$ and $R^o_\Sigma(\tau)$, of the corresponding autocorrelation functions are shown in Fig.\,12\,c, when $\gamma=\pi$, and in Fig.\,12\,d, when $\gamma=3\mspace{1mu}\pi$. In both the cases, the value of compression gain, 
$G_C$, is very close to $G_C=2$.

\subsubsection{Hermite-Gaussian Pulse}
In this case, the antisymmetric component, $y(t)$, of the pulse (25) is given by (23), and its 
autocorrelation function is of the form
\begin{equation}
	R_{yy}(\tau) \,=\, \sqrt{\pi}\mspace{2mu} 
	(2-\lambda^2\tau^2\mspace{1mu}) \exp\mspace{1mu}(-\lambda^2\tau^2\!/4)/(4\mspace{1mu}\lambda^3)
\end{equation}
where $\lambda>0$ is the scale parameter.

The autocorrelation function, $R_\Sigma(\tau)$, determined from (33), can be expressed as
\begin{equation}
	R_\Sigma(\tau) \,=\,G_{\mspace{-2mu} \lambda} \mspace{1mu}
	\big[4\mspace{1mu}{\Gamma}(3/2;\lambda^2\tau^2\!/4) -
	\lambda^2\tau^2 \mspace{1mu}{\Gamma}(1/2;\lambda^2\tau^2\!/4)\big] \,
\end{equation}
where $G_{\mspace{-2mu} \lambda} \,\triangleq\, \exp\mspace{1mu}(-\lambda^2\tau^2\!/4) / (2\mspace{1mu}\lambda^3)$, and
\begin{equation*}
	\Gamma(\nu;\zeta) \,\triangleq \int_\zeta^\infty \mspace{-5mu} \rho^{\mspace{1mu}\nu-1}\mspace{-2mu}\exp(-\rho)\mspace{1mu} d\rho
\end{equation*}
is the incomplete gamma function [7, (8.2.2)].

By using the relationships [10],
\begin{eqnarray*}
	\Gamma(1/2;\zeta^2)\!\! &\!\!=\!\!& \! \sqrt{\pi}\, \mathrm{erfc}(\zeta) \\
	\Gamma(\nu\!+\!1;\zeta)\!\! &\!\!=\!\!& \nu\mspace{1mu}\Gamma(\nu;\zeta) + \zeta^\nu \mspace{-3mu}\exp(-\zeta)
\end{eqnarray*}
the autocorrelation function, $R_\Sigma(\tau)$, can be expressed in an equivalent, 
yet less compact, form
\begin{eqnarray*}
	R_\Sigma(\tau)\, =   \mspace{-25mu}  &{}&\big[1/(2\lambda^3)\big] \Big [ \mspace{-2mu}  
	2\lambda|\tau| \exp\mspace{1mu}(-\lambda^2\tau^2\!/2) \\
	&{}& \mspace{-14mu}-\,\sqrt{\pi} \mspace{2mu} (\lambda^2\tau^2\mspace{-6mu}-\!2) 
	\exp\mspace{1mu}(-\lambda^2\tau^2\!/4)\mspace{2mu}
	\mathrm{erfc}\mspace{1mu}\big(\lambda|\tau|/2
		\big) \mspace{-2mu}	\Big ].
\end{eqnarray*}

The autocorrelation function, $R_{xx}(\tau)$, of the symmetric component, $x(t)$, of the pulse (25) is equal to the difference$\mspace{1mu}${\footnote {
\,$R_{xx}(\tau)$ can also be determined directly from (67).}}, $[R_\Sigma(\tau)\!-\!R_{yy}(\tau)]$. 
In particular, when $\lambda^2=1/2$,
\begin{eqnarray*}
	R^{\mspace{1mu}o}_{xx}(\tau) \!\! &\!\!=\!\!& \\
	 &\!\!{}\!\!& \!\!\!\!	\big[ 16\mspace{1mu}{\Gamma}(3/2;\tau^2\!/8)-
	2\mspace{1mu}\tau^2 \mspace{1mu}{\Gamma}(1/2;\tau^2\!/8)
	+ \sqrt{\pi}(\tau^2\!-\!4)
	\big]\\
	 &\!\!{}\!\!& \!\!\!\!\times \exp(-\tau^2\!/8) /(4\sqrt{\pi}).
\end{eqnarray*}
The characteristic shape of the resulting function $R_{xx}(\tau)$ is typical for a broad class of positive bimodal pulses.

Normalised forms, $R^o_{xx}(\tau)$, $R^o_{yy}(\tau)$ and $R^o_\Sigma(\tau)$, of the corresponding autocorrelation functions are shown in Fig.\,12\,e. In this case, the compression gain $G_C \approx 2.3$.

\subsection{Quadrature Pulse Compressor}
A block diagram of a quadrature receiver/pulse compressor is depicted in Fig.\,13. The system consists of two parallel channels; each channel comprises a multiplier 
$\mspace{-2mu}{\boldsymbol{\times}}$, followed by a low-pass filter {\sf{\small{LPF}}} 
and a bank of two parallel filters, {\sf{\small{MFX}}} and {\sf{\small{MFY}}}, matched to waveforms, $x(t)$ and $y(t)$, respectively. 

In the upper channel, the reference input of the multipler is 
driven by a cosinusoidal signal 
$2\mspace{-1mu} \cos \mspace{-2mu}\left(\mspace{1mu} \omega_0\mspace{1mu}t \!+\! \phi \right)$,
 whereas in the lower channel, a sinusoidal signal 
$2\mspace{-1mu} \sin \mspace{-2mu}\left(\mspace{1mu} \omega_0\mspace{1mu}t \!+\! \phi \right)$
is used as a reference. 
The phase angle $\phi$ represents a circular sum of the phase offset between transmitter and receiver, and also the phase shift resulting from the transmission path.

In general, an input signal, $r(t)$, to be processed is an attenuated version of a transmitted signal, $s(t)$, corrupted by additive noise. However, for the purpose of this analysis it is assumed that 
\begin{equation}
r(t)\, \equiv \, s(t) \,=\, 
x(t)\cos(\omega_0\mspace{1mu} t) + y(t)\sin(\omega_0\mspace{1mu} t).	
\end{equation}
Consequently, the output signals, $u(t)$ and $v(t)$, of the respective low-pass filters can be expressed as
\begin{eqnarray}
	u(t) \!\! &\!\!=\!\!& \! x(t) \cos(\phi) -y(t) \sin(\phi)\\
	v(t) \!\! &\!\!=\!\!& \! x(t) \sin(\phi) +y(t) \cos(\phi).
\end{eqnarray} 

Each of the above signals is applied to the bank of two parallel matched filters, {\sf{\small{MFX}}} and {\sf{\small{MFY}}}. The four ouputs, $a(\tau), b(\tau), c(\tau)$ and $d(\tau)$, of the matched filters are given by
\begin{eqnarray}
	a(\tau) \!\! &\!\!=\!\!& \! R_{xx}(\tau) \cos(\phi) -R_{yx}(\tau) \sin(\phi)
	\nonumber \\
	b(\tau) \!\! &\!\!=\!\!& \! R_{xy}(\tau) \cos(\phi) -R_{yy}(\tau) \sin(\phi)
	\nonumber \\
	c(\tau) \!\! &\!\!=\!\!& \! R_{xx}(\tau) \sin(\phi) +R_{yx}(\tau) \cos(\phi)
	\nonumber \\
	d(\tau) \!\! &\!\!=\!\!& \! R_{xy}(\tau) \sin(\phi) +R_{yy}(\tau) \cos(\phi)
\end{eqnarray} 
where the two cross-correlation functions are defined by
\begin{eqnarray*}
	R_{xy}(\tau) \! &\!\!\triangleq\!\!& \!\!
	\int_{-\infty}^{\infty}\!\!
	x(t)\mspace{1mu} y(t+\tau) \, dt\\
	R_{yx}(\tau) \! &\!\!\triangleq\!\!& \!\!
	\int_{-\infty}^{\infty}\!\!
	y(t)\mspace{1mu} x(t+\tau) \, dt.
\end{eqnarray*}
Adding the two conditions $\sin^2(\phi)+\cos^2(\phi)=1$ and $R_{yx}(\tau)= -R_{xy}(\tau)$ (see Appendix)
makes it possible, in principle, to determine all unknown quantities of interest. However, in the presence of noise, a more robust solution will be obtained, when a phase-invariant procedure is applied.

Consider the following correlation matrix
\begin{equation*}
	\mathbf{M}_{xy}(\tau) =
	\left [
	\begin{array}{r@{\quad}l}\!\!a(\tau) & b(\tau)\!\\
		\!\!c(\tau) & d(\tau)\!
			\end{array} \right ] .  
\end{equation*}
The determinant $\mathcal{D}_{\mathbf{M}}(\tau)$ of $\mathbf{M}_{xy}(\tau)$ can be expressed as
\begin{eqnarray}
	\mathcal{D}_{\mathbf{M}}(\tau) &\!\!\triangleq\!\!& a(\tau)\mspace{1mu}d(\tau) - b(\tau)\mspace{1mu}c(\tau) \nonumber\\
	&\!\!=\!\!& R_{xx}(\tau)\mspace{1mu} R_{yy}(\tau)  -  R_{xy}(\tau) \mspace{1mu} R_{yx}(\tau)
	\end{eqnarray} 
so that (see Appendix) 
\begin{equation}
	\mathcal{D}_{\mathbf{M}}(\tau) \,=\, [R_{xx}(\tau)+ R_{yy}(\tau)]^2\mspace{-1mu}/4
	\,=\, R^2_\Sigma(\tau)/4.
\end{equation} 
As seen, the determinant $\mathcal{D}_{\mathbf{M}}(\tau)$ is phase 
invariant$\mspace{1mu}${\footnote {\,The function, $a^2(\tau)\!+\!b^2(\tau)\!+\!c^2(\tau)\!+\!d^2(\tau)$, is also phase invariant.}}.

The quadrature receiver/pulse compressor, shown in Fig.\,13, employs a low-complexity
 signal processor performing the following operation
\begin{eqnarray}
&{}&\mspace{-22mu}[a(\tau)\!+\!d(\tau)]^2 - [a(\tau)\!-\!d(\tau)]^2 - 
	[b(\tau)\!+\!c(\tau)]^2 \nonumber\\	
+\mspace{-30mu}&{}&   [b(\tau)\!-\!c(\tau)]^2
		\,=\, 4\mspace{1mu}\mathcal{D}_{\mathbf{M}}(\tau) \,=\, 
		R^2_\Sigma(\tau).\qquad \quad
\end{eqnarray}

\begin{figure}[]%[b]
	\centering
	\includegraphics[width=8.5cm]{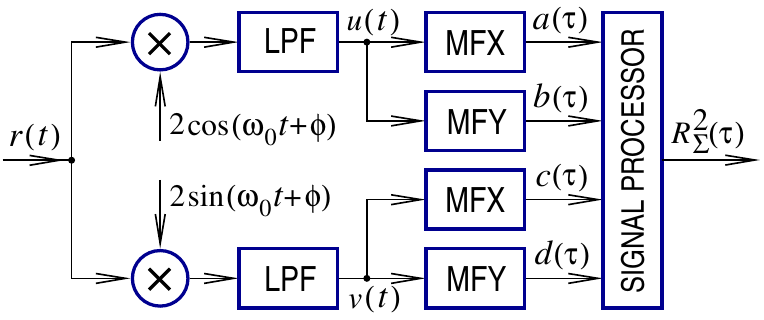}
	\vspace{-0.2cm}
	\caption{Quadrature receiver/pulse compressor.}
\end{figure}

\vspace{0.05cm}
\noindent {\emph{Example 4:}} To elucidate the relationship between the determinant, $\mathcal{D}_{\mathbf{M}}(\tau)$, and the correlation function, $R_\Sigma(\tau)$, consider the case of a minimally compressible Laplacian pulse (41).

From (46) and (47), it follows that
\begin{equation*}
R_{xx}(\tau)\mspace{1mu}R_{yy}(\tau) \,=\,
	(1-\beta^2 \mspace{1mu}|\tau|^2)\exp(-2\mspace{1mu}\beta\mspace{1mu}|\tau|)/\beta^2.
\end{equation*}
Then, according to Wiener-Khinchin theorem, the two crosscorrelation functions, $R_{xy}(\tau)$ and $R_{yx}(\tau)$, can be determined from
\begin{equation*}
	R_{xy}(\tau) \!=\! \mathcal{F}^{-1}\{X^*(\omega)Y(\omega)\}, \mspace{11mu}
	R_{yx}(\tau) \!=\! \mathcal{F}^{-1}\{X(\omega)Y^*(\omega)\} \nonumber 
\end{equation*}
where $X(\omega)$ and $Y(\omega)$ are given by (42).\\
Therefore,
\begin{eqnarray*}
	R_{xy}(\tau)\,R_{yx}(\tau)&\!\!=\!\!& 
	-\mspace{-2mu}\left[\mathcal{F}^{-1}\{4\mspace{2mu} \beta \omega/(\omega^2+\beta^2)^2\}
	\right]^2 \nonumber\\
	&\!\!=\!\!& 
	-\, |\tau|^2 \exp(-2\mspace{1mu}\beta\mspace{1mu}|\tau|).	\qquad
\end{eqnarray*} 

Consequently, the determinant $\mathcal{D}_{\mathbf{M}}(\tau)$ 
can be expressed as follows
\begin{eqnarray*}
	\mathcal{D}_{\mathbf{M}}(\tau) &\!\!=\!\!& 
	R_{xx}(\tau)\mspace{1mu} R_{yy}(\tau)-R_{xy}(\tau) \mspace{1mu} R_{yx}(\tau)
	\nonumber\\
	&\!\!=\!\!& \exp(-2\mspace{1mu} \beta |\tau|)/\beta^2.
\end{eqnarray*} 
Then, since from (43), $R_\Sigma(\tau)= 2\mspace{1mu} \exp(-\beta |\tau|)/\beta$,
\begin{equation*}
	\mathcal{D}_{\mathbf{M}}(\tau) \,=\, R^2_\Sigma(\tau)/4
\end{equation*} 
in agreement with (61).

\section{Conclusions}
The above analysis has shown that the achievable compression gain $G_C$ depends on the shape of a minimally compressible pulse. In the case of a unimodal pulses, the gain decreases from 
$G_C \approx 2.42$, for a Laplacian (leptokurtic) shape to $G_C=2$, for a rectangular (platykurtic) shape.

In the case of a bimodal Hermite-Gaussian pulse, the achievable compression gain, 
$G_C \approx 2.3$, is greater than that, $G_C \approx 2.1$, obtained for a unimodal Gaussian 
(mesokurtic) shape.

Although the compression gain is small, the generation and phase-invariant correlation processing of such minimally-compressible pulses can easily be implemented with standard low-complexity analogue or digital building blocks.

The concept of phase-invariant correlation processing can be extended to include other waveforms of 'complementary' nature, such as asynchronous binary waveforms  and also frequency modulated signals.

\section*{Acknowledgment}
The author is grateful to Prof. Miroslaw Bober for the many stimulating discussions on unconventional approaches to signal and image processing.

\section*{Appendix\\	Correlation Functions}
\noindent  {\emph{Unimodal Case:}} Assume that $\tau \ge 0$ and define two auxiliary functions, $\mathsf{A}(\tau)$ and $\mathsf{B}(\tau)$, as follows (see Fig.\,14\,a)
\begin{eqnarray}
	\mathsf{A}(\tau) \mspace{-8mu}&\triangleq&\mspace{-8mu} \int_{-\infty}^{-\tau} \mspace{-5mu}
	x(t)\mspace{1mu} x(t+\tau) \, dt \,\,
	= \int_{\mspace{-1mu}0}^{\mspace{1mu}\infty} \mspace{-5mu}
	x(t)\mspace{1mu} x(t+\tau) \, dt \nonumber \quad \\
	\mathsf{B}(\tau) \mspace{-8mu}&\triangleq&\mspace{-8mu} \int_{-\tau}^{\mspace{1mu}0} \mspace{-5mu}
	x(t)\mspace{1mu} x(t+\tau) \, dt.
	\nonumber 
\end{eqnarray} 
Accordingly, the autocorrelation functions, $R_{xx}(\tau)$ and $R_{yy}(\tau)$, can be expressed as
\begin{eqnarray}
	R_{xx}(\tau)\!\! &\!\!=\!\!& \! 
	\int_{-\infty}^{\infty}\!\!
	x(t)\mspace{1mu} x(t+\tau) \, dt 	\nonumber\\
	&\!\!=\!\!& 2\mspace{1mu} \mathsf{A}(|\tau|) + \mathsf{B}(|\tau|)\nonumber\\
	R_{yy}(\tau)\!\! &\!\!=\!\!& \! 
		\int_{-\infty}^{\infty}\!\!
	x(t)\mspace{1mu} x(t+\tau)\mspace{1mu}
	\mathrm{sgn}(t)\mspace{1mu} \mathrm{sgn}(t+\tau) \, dt \nonumber\\
	&\!\!=\!\!& 2\mspace{1mu} \mathsf{A}(|\tau|) - \mathsf{B}(|\tau|).\nonumber
\end{eqnarray}
Consequently, the real part, $R_\Sigma(\tau)$, of the complex autocorrelation function $R_{zz}$ can be determined from
\begin{eqnarray}
	R_\Sigma(\tau)\! &\!\!\triangleq\!\!& \!
	\mathcal{R}e\{R_{zz}(\tau)\}
	= R_{xx}(\tau)+ R_{yy}(\tau)
	\nonumber\\
	\!\! &\!\!=\!\!& \! 4\mspace{1mu} \mathsf{A}(|\tau|) =
		 4\mspace{-4mu} 	\int_{\mspace{-1mu}0}^{\infty} \mspace{-5mu}
	x(t)\mspace{1mu} x(t+|\tau|) \, dt. 
\end{eqnarray}

The two cross-correlation functions, $R_{xy}(\tau)$ and $R_{yx}(\tau)$, can be expressed as
\begin{eqnarray}
	R_{xy}(\tau)\!\! &\!\!=\!\!& \! 
	\int_{-\infty}^{\infty}\!\!
	x(t)\mspace{1mu} x(t+\tau)\mspace{1mu}
	 \mathrm{sgn}(t+\tau) \, dt\nonumber\\
	&\!\!=\!\!& \mathsf{B}(|\tau|)\,\mathrm{sgn}(\tau) \nonumber\\
	R_{yx}(\tau)\!\! &\!\!=\!\!& \! 
	\int_{-\infty}^{\infty}\!\!
	x(t)\mspace{1mu} x(t+\tau)\mspace{1mu}
	\mathrm{sgn}(t) \, dt\nonumber\\
	&\!\!=\!\!&  - \mathsf{B}(|\tau|)\,\mathrm{sgn}(\tau)\nonumber.
\end{eqnarray}
Therefore, the imaginary part, $R_\Delta(\tau)$, of the complex autocorrelation function $R_{zz}$ can be determined from
\begin{eqnarray}
	R_\Delta(\tau)\! &\!\!\triangleq\!\!& \!  
	\mathcal{I}m\{R_{zz}(\tau)\}
	= R_{xy}(\tau)- R_{yx}(\tau)\nonumber\\
	\!\! &\!\!=\!\!& \! 2\mspace{3mu}\mathrm{sgn}(\tau) \mspace{1mu} \mathsf{B}(|\tau|)\\
	 \!\! &\!\!=\!\!& \!
	2\mspace{3mu} \mathrm{sgn}(\tau) \mspace{-4mu}
	\int_{-|\tau|}^{\mspace{1mu}0} \mspace{-5mu}
	x(t)\mspace{1mu} x(t+|\tau|) \, dt.
\end{eqnarray} 
The imaginary part can also be expressed as
\begin{equation*}
	R_\Delta(\tau) \, = \,  
	\left[R_{xx}(\tau)\!-\! R_{yy}(\tau)\right]\mathrm{sgn}(\tau).
\end{equation*} 

The determinant (60) of the correlation matrix $\mathbf{M}_{xy}(\tau)$ can be expressed as
\begin{eqnarray}
	\mathcal{D}_{\mathbf{M}}(\tau) &\!\!\!=\!\!& R_{xx}(\tau)\mspace{1mu} R_{yy}(\tau)  -  R_{xy}(\tau) \mspace{1mu} R_{yx}(\tau)
	\nonumber \\
	&\!\!\!=\!\!&  (2\mspace{1mu} \mathsf{A}(|\tau|) - \mathsf{B}(|\tau|))(2\mspace{1mu} \mathsf{A}(|\tau|)- \mathsf{B}(|\tau|)) + \mathsf{B}^2(|\tau|)\nonumber \\ 
	&\!\!\!=\!\!& 4\,\mathsf{A}^2(|\tau|)  = R^2_\Sigma(\tau)/4.
\end{eqnarray}

\begin{figure}[] %[b]
	%\vspace{-0.1cm}
	\centering
	\includegraphics[width=8.1cm]{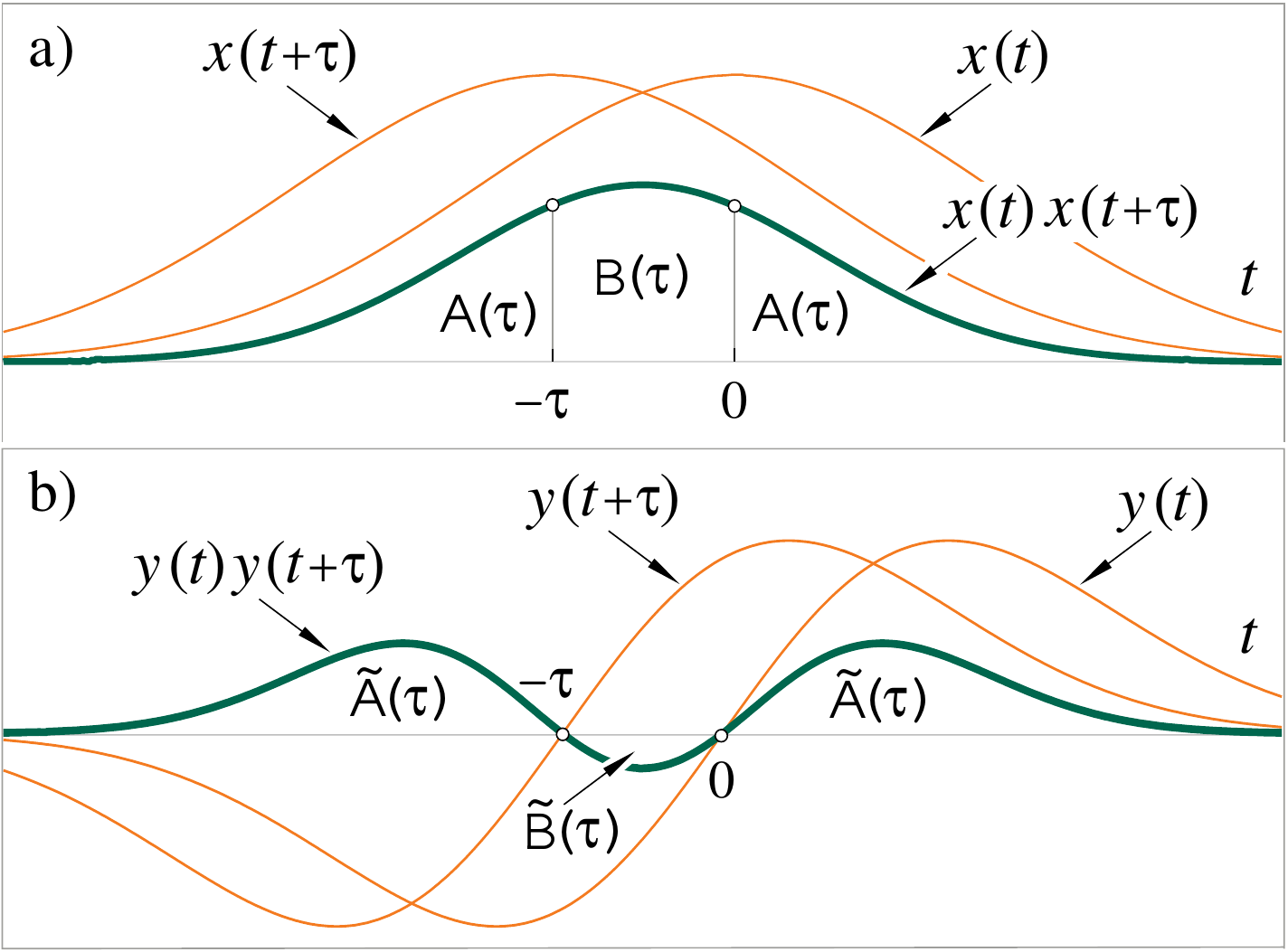}
	\vspace{-0.26cm}
	\caption{Representation of two auxiliary functions $\mathsf{A}(\tau)$ and $\mathsf{B}(\tau)$: unimodal case (a) and bimodal case (b).}
\end{figure}

\vspace{0.1cm}
\noindent  {\emph{Bimodal Case:}} Assume that $\tau \ge 0\mspace{1mu}$; 
two auxiliary functions, $\widetilde{\mathsf{A}}(\tau)$ and $\widetilde{\mathsf{B}}(\tau)$, 
are defined by (see Fig.\,14\,b)
\begin{eqnarray*}
	\widetilde{\mathsf{A}}(\tau) \mspace{-8mu}&\triangleq&\mspace{-8mu} \int_{-\infty}^{-\tau} \mspace{-5mu}
	y(t)\mspace{1mu} y(t+\tau) \, dt \,\,
	= \int_{\mspace{-1mu}0}^{\mspace{1mu}\infty} \mspace{-5mu}
	y(t)\mspace{1mu} y(t+\tau) \, dt  \quad \\
	\widetilde{\mathsf{B}}(\tau) \mspace{-8mu}&\triangleq&\mspace{-8mu} \int_{-\tau}^{\mspace{1mu}0} \mspace{-5mu}
	y(t)\mspace{1mu} y(t+\tau) \, dt.
\end{eqnarray*}
Therefore,
${}$

\vspace{-0.7cm}
\begin{eqnarray}
R_{xx}(\tau)\!\! &\!\!=\!\!& \! 2\mspace{1mu} \widetilde{\mathsf{A}}(|\tau|) 
+ \big|\widetilde{\mathsf{B}}(|\tau|)\big|\\
R_{yy}(\tau)\!\! &\!\!=\!\!& \! 2\mspace{1mu}\widetilde{\mathsf{A}}(|\tau|) 
- \big|\widetilde{\mathsf{B}}(|\tau|)\big|
\end{eqnarray}
so that $R_\Sigma(\tau) = 4\mspace{1mu} \widetilde{\mathsf{A}}(|\tau|)$ and 
$R_\Delta(\tau) = 2\mspace{3mu}\mathrm{sgn}(\tau) \mspace{1mu}\big| \widetilde{\mathsf{B}}(|\tau|)\big|$.

\end{document}